\documentclass[floatfix,twocolumn,superscriptaddress,preprintnumbers,amsmath,amssymb,pra]{revtex4-2}
\pdfoutput=1
\usepackage{amsmath}
\usepackage{graphicx}
\usepackage{dcolumn}
\usepackage{appendix}
\usepackage{bm}
\usepackage{comment}
\usepackage{chngcntr}
\usepackage{multirow}
\usepackage{booktabs}
\usepackage{subfigure}
\usepackage{xcolor}
\usepackage[colorlinks,linkcolor=blue,anchorcolor=blue,citecolor=blue]{hyperref}
\usepackage[capitalise]{cleveref}
\usepackage[normalem]{ulem}

\begin{document}
\title{Response of asymmetric nuclear matter studied with a finite number of particles}
\date{\today}

\newcommand{\FM}[1]{{\color{magenta} #1}}
\newcommand{\gc}[1]{{\color{blue} #1}}
\newcommand{\ML}[1]{{\color{olive} #1}}

\author{M. Li}
\affiliation{Department of Physics, School of Science, The University of Tokyo, Tokyo 113-0033, Japan}
\affiliation{Quark Nuclear Science Institute, The University of Tokyo, Tokyo 113-0033, Japan}
\author{F. Marino}
\affiliation{Institut f\"{u}r Kernphysik and PRISMA+ Cluster of Excellence, Johannes Gutenberg-Universit\"{a}t Mainz, 55128 Mainz, Germany}
\author{G. Col\`o}
\affiliation{Dipartimento di Fisica “Aldo Pontremoli,” Universit\`a degli Studi di Milano, 20133 Milano, Italy}
\affiliation{Istituto Nazionale di Fisica Nucleare, Sezione di Milano, 20133 Milano, Italy}
\author{H. Liang}
\affiliation{Department of Physics, School of Science, The University of Tokyo, Tokyo 113-0033, Japan}
\affiliation{Quark Nuclear Science Institute, The University of Tokyo, Tokyo 113-0033, Japan}
\affiliation{Interdisciplinary Theoretical and Mathematical Sciences Program (iTHEMS), RIKEN, Wako, Saitama 351-0198, Japan}

\begin{abstract}
We investigate the ground-state properties of asymmetric nuclear matter and its response to a static perturbation using the density functional theory framework.
Our method, which extends the finite-nucleon-number technique of [Phys. Rev C 107, 044311 (2023)] to the case of isospin-asymmetric matter, allows to study the impact of external isoscalar and isovector fields on the system.
In particular, the densities and static response functions in the different isospin channels, and as a function of isospin asymmetry, are evaluated.
Finite-size effects are discussed by comparing with Random Phase Approximation predictions in the thermodynamic limit.
A peculiar non-monotonic behavior of the isovector response function is analyzed.
\end{abstract}

\maketitle

\section{Introduction} \label{intro}
The study of nuclear structure, which focuses on various properties of nuclei, involves solving the quantum many-body problem.
Among the different theoretical approaches, the \textit{ab initio} methods~\cite{10.3389/fphy.2020.00379,10.3389/fphy.2023.1129094,papenbrock2024}, which solve the many-nucleon Schr\"odinger equation starting from realistic nuclear interactions,
aim to provide systematically improvable predictions with controllable theory uncertainties.
However, the \textit{ab initio} methods are very complicated and highly demanding in their computational cost.
Therefore, the \textit{ab initio} methods are limited to relatively light nuclei ($A \approx 140$) with closed shell, although they have gone through a dramatic development in recent years~\cite{10.3389/fphy.2020.00379,PbAbInitio}.

To perform global calculations in heavy nuclei, and to excited states, the density functional theory (DFT) \cite{Kohn_1965,doi:10.1080/23746149.2020.1740061,AFANASJEV2013680,Kortelainen2014} 
is the only microscopic choice nowadays and even in the near future.
Indeed, DFT can routinely deal with almost all the nuclei in the nuclear chart, with the exception of some very light nuclei, at a relatively mild computational cost.
In DFT, the interacting nuclear systems are solved by mapping them to an auxiliary problem of non-interacting particles, such that the same ground state densities of the true system are obtained.
Hence, the many-body problem is recast into self-consistent single-particle (s.p.) problem. 
DFT is, in principle, an exact theory; however, its basic object, the energy as a function of the density $E[\rho]$, which is dubbed energy density functional or EDF, is not known exactly. 
It is non-trivial to devise accurate EDFs, especially in the case of nuclear systems.
Most existing studies of nuclear DFT construct the EDF empirically, i.e., some ansatz for the form of the EDF is taken as a starting point, and the parameters are fitted to observables such as experimental data, typically masses and radii of stable nuclei close to magicity~\cite{doi:10.1080/23746149.2020.1740061}. 
Empirical EDFs are successful in reproducing the experimental data for many more nuclei than those used in the fit, but show increasingly large uncertainties outside the regions where they have been tuned, e.g., in nuclei far from stability~\cite{AFANASJEV2013680} or with large isospin asymmetries.
Furthermore, over the last decade, it has been found hard to dramatically improve the quality of empirical EDFs, see e.g.~Ref.~\cite{Kortelainen2014}.

The limitations of the empirical EDFs have motivated investigations into constructing EDFs grounded in \textit{ab initio} theory, see e.g.~\cite{Drut2010,Duguet2023,Zurek:2023mdh,MarinoPhdThesis,Marino2021,Reed:2025ccn} and references therein.
The strategy explored in Refs.~\cite{Marino2021,Marino2023,MarinoPhdThesis,ColoQnp} takes inspiration from the ``Jacob's ladder''~\cite{10.1063/1.1390175,Martin_2004} approach of electronic DFT, which focuses on the construction of a hierarchy of EDFs of increasing complexity and accuracy.
The first ''rung'' is the local density approximation (LDA), where the EDF solely depends on the number density of the system.
The LDA EDF is directly obtained from the equation of state (EoS) of uniform matter (the homogeneous electron gas in condensed matter, infinite nuclear matter in the context of nuclear physics), as obtained from accurate \textit{ab initio} techniques.
The nuclear matter EoS is actively studied for its relevance in describing neutron stars and astrophysical scenarios (see e.g.~\cite{NeutronStarBook,BURGIO2021103879,NeutronMatter2024,margueron2025nucleardatapytoolkitsimpleaccess}), as well as for its connections to finite nuclei~\cite{rocamaza2018}.
Among the \textit{ab initio} methods for studying nuclear matter, we mention the Brueckner-Hartree-Fock method \cite{PhysRevC.75.025802, doi:10.1142/9789812799760_0008}, self-consistent Green's functions theory in the finite-temperature~\cite{Barbieri2017,Rios2020,Carbone2020} and algebraic diagrammatic construction~\cite{Barbieri2017,Marino2024,Marino2025Qnp,MarinoAdc} schemes, coupled-cluster theory~\cite{Hagen2014,Jiang2024}, and quantum Monte Carlo (QMC)~\cite{RevModPhys.87.1067,Gandolfi2015,Lynn2019,Tews2020,Roggero2014CIMC}.

However, LDA is insufficient for describing nuclei, as surface effects are missing in homogeneous matter.
Surface terms, such as density gradients and spin-orbit contributions, must be constrained in inhomogeneous systems.
This has motivated attempts to extract the surface terms from neutron and neutron-proton drops, i.e., nuclear matter bound in a trap studied \textit{ab initio} (see e.g.~\cite{PhysRevLett.76.2416,PhysRevLett.106.012501,Shen2018,SHEN2019PPNP}).
Moreover, the possibility of using the static response of nuclear matter perturbed by a weak external potential as a way of providing constraints to the surface part of the EDF has been investigated in Refs.~\cite{carlson2014,Gezerlis2017,Gezerlis2021,MarinoPhdThesis,ColoQnp} in conjunction with QMC computations.

Perturbed matter is an appealing model system, since, for weak enough external potentials, linear response theory~\cite{fetter2003quantum,Martin_2004} provides a clear connection between pseudo-observables, such as the energy per particle or the amplitude of the density fluctuations, and the strength and momentum of the external field.
This also allows, in principle, for a well-defined way of mapping the \textit{ab initio}, ``exact'' response to the effective EDF description, as demonstrated by calculations in electronic systems~\cite{Dornheim2018Review,MOLDABEKOV2025104144} and cold atoms~\cite{carlson2014}.
While determining the static response is a numerically delicate problem~\cite{ColoQnp,MarinoPhdThesis}, developments in addressing response functions calculated with \textit{ab initio} promise that sufficient accuracy may be reached in the near future~\cite{Gezerlis2021,Sobczyk2025}.

Since most \textit{ab initio} methods perform calculations for nuclear matter with a finite number of particles, a DFT method to deal with a finite number of nucleons has been developed in Refs.~\cite{Gezerlis2022Skyrme,Marino2023}: the nucleons are confined in a box subject to periodic boundary conditions (PBCs) to simulate infinite matter.
However, the method in Ref.~\cite{Marino2023} is limited to symmetric nuclear matter (SNM) and pure neutron matter (PNM). 
The response of asymmetric nuclear matter (ANM) has not been considered and, also, the difference between the response to isoscalar and isovector operators has not been investigated. 
Such extensions are called for, to make the tool appropriate for a mapping of {\it ab initio} studies.

In fact, nuclear interactions are strongly isospin-dependent, and so are the EDFs~\cite{CHABANAT1997710}.
In addition, most nuclei on the nuclear chart have different numbers of protons and neutrons and an increasing number of neutron-rich or neutron-deficient nuclei are being discovered.
Existing EDFs are fit mostly to nuclei close to the stability valley. 
Hence, the isovector contributions of the EDF are somewhat poorly constrained, which impairs the predictive power of EDFs far from stability.
Finally, having a sound, {\em ab initio}-based EDF for neutron-rich nuclei and ANM is relevant for
advancing our understanding of compact objects, in particular of the crust and the outer core of
neutron stars~\cite{BURGIO2021103879}

Consequently, the extension to ANM and isoscalar/isovector operators
is an important step in the project started with Ref.~\cite{Marino2021}, which focuses on the construction of \textit{ab initio}-based nuclear EDFs.
In this work, the finite-nucleon DFT method is extended to the ANM.
Understanding FS effects is of paramount importance to properly map the information in the \textit{ab initio} finite-$A$ computations to the EDFs.
Thus, we have investigated the FS effects in asymmetric nuclear matter both in the homogeneous phase and in the system subject to an external field.
This work provides guidance for fitting the nuclear EDF to the \textit{ab initio} results in isospin-asymmetric matter.

This paper is organized as follows. 
In \cref{secTF}, we summarize the finite-nucleon DFT method and the theory of static response, and extend the approach to ANM.
In Sec.~\ref{NT}, we provide a numerical validation of our technique.
Results concerning perturbed ANM are then shown in \cref{results}.
Finally, conclusions and perspectives are drawn in \cref{Conclusion}.

\section{Theoretical formalism}\label{secTF}

\subsection{Asymmetric nuclear matter with periodic boundary conditions}\label{ANM}

In this work, a DFT method to study asymmetric nuclear matter (ANM) in the presence of an external field is detailed, where periodic boundary conditions (PBCs) are employed to simulate the infinite system.
In particular, we concentrate on the zero-temperature and spin-saturated matter.

Nuclear matter is an idealized extended system containing an infinitely large number of nucleons $A$ in an infinite volume $V$, such that the density $\rho_0=A/V$ is homogeneous and finite~\cite{FIORILLA2012317,Margueron2018,rocamaza2018}.
In our framework, the particles interact only through the strong interaction, while the Coulomb force is neglected.
Isospin-asymmetric nuclear matter consists of different numbers of neutrons and protons.
The system is then characterized by the isospin asymmetry $\beta = \rho_1/\rho_0$, where $\rho_0 =\rho_n + \rho_p$ and $\rho_1 = \rho_n - \rho_p$ are the isoscalar and isovector number densities, respectively.

Nuclear matter can be studied with a variety of techniques, including DFT and \textit{ab initio} methods.
Most of the latter employ a finite number of nucleons to simulate nuclear matter (see e.g.~\cite{Hjorth-Jensen:2017gss,Hagen2014,Marino2024,Tews2020}), and the same approach will be followed in this work, extending the finite-$A$ DFT scheme of Refs.~\cite{Gezerlis2022Skyrme,Marino2023}.
In this approach, a finite number of particles $A$ are put in a cubic box, whose size $L$ and volume $V=L^3$ are determined from the density $\rho_0$ of the system, i.e., $V = L^3 = A/\rho_0$.
The thermodynamic limit (TL) is obtained by letting both the number of particles $A$ and the box size $L$ go to infinity while simultaneously keeping the density $\rho_0$ fixed \cite{Hjorth-Jensen:2017gss}.

In a uniform system, the s.p. wave functions are plane waves,
\begin{align}
\psi_{\mathbf{k}}(\mathbf{x}) = \frac{e^{i\mathbf{k} \cdot \mathbf{x}}}{\sqrt{V}}, \label{planewave}
\end{align}
where $\mathbf{k}$ and $\mathbf{x}$ are the momentum and space coordinates, respectively.
Thus, the s.p. energy solely depends on the square of the momentum $\mathbf{k}^2$, which, as a consequence of PBCs, turns out to be quantized as $\mathbf{k} = \frac{2\pi}{L} \mathbf{n}$, where $\mathbf{n} = (n_x,n_y,n_z)$ is a vector of three integer numbers.
Consequently, the system shows a shell structure in momentum space and the ``magic numbers'', corresponding to filling completely a certain number of momentum shells, read $A/g = 1$ when states up to $n^2 = n_x^2 + n_y^2 + n_z^2 = 0$ are occupied, 7 for $\mathbf{n}^2 \leq 1$, 19 for $\mathbf{n}^2 \leq 2$, 27 for $\mathbf{n}^2 \leq 3$, 33 for $\mathbf{n}^2 \leq 4$, etc., where $A$ is the number of nucleons (see e.g.~\cite{Marino2023,LietzCompNucl}).

In ANM, $g=2$ accounts for the spin-projection degeneracy.
The same shell structure shows up for both neutrons and protons.
To ensure that the density at the ground state is uniform, the number of particles should be chosen as one of the magic numbers.

\subsection{Nuclear DFT description of nuclear matter} \label{DFT}

In this section, we will first outline the fundamentals of nuclear DFT \cite{Kohn_1965, Schunk2019, Colo01012020} and then specialize the formalism to the case of interest.
In previous works \cite{Marino2023,Gezerlis2022Skyrme,MarinoPhdThesis}, a DFT method using PBCs was established for SNM and PNM.
In this work, this method is extended to the case of spin-saturated ANM, and external perturbations in both the isoscalar and isovector channels are considered.
Isoscalar quantities are related to overall properties of the system, while isovector quantities, account for the difference of neutron and proton properties.
Examples of isovector properties include the nuclear symmetry energy and, in finite nuclei, the thickness of the neutron skins~\cite{rocamaza2018,Piekarewicz2019NeutronSkin}.

In the Kohn-Sham (KS) formulations of DFT~\cite{Colo01012020,Giuliani_Vignale_2005}, the g.s. wavefunction of an interacting system is replaced by a Slater determinant of effective s.p. wave functions called the KS orbitals, defined such that the g.s. density generated by those KS orbitals matches the true density of the interacting system \cite{Kohn_1965}.
The KS equations have the form of s.p. Schr\"{o}dinger equations, where particles move in a self-consistent potential, which can be conveniently derived from an EDF,  i.e. an expression for the total energy as a functional of the densities of the system.

The EDF for a system in the presence of an external perturbation is given by
\begin{align}
E[v,\rho]= E_{\rm kin}+E_{\rm pot}+E_{\rm ext},\label{totalenergy}
\end{align}
where $E_{\rm kin}$ is the non-interacting kinetic energy, $E_{\rm pot}$ the potential energy, and $E_{\rm ext}$ the contribution of the external potential.
For later convenience, the total energy is expressed as a functional of both the densities $\rho$ and the external field $v$.
The potential energy is given by 
\begin{align}
E_{\rm pot}=\int d\mathbf{x} \, \mathcal{E}_{\rm pot}(\mathbf{x})\label{pot},
\end{align}
where $\mathcal{E}_{\rm pot}(\mathbf{x})$ denotes the potential energy density.
Moreover, the kinetic energy and the energy of the external field are calculated with
\begin{align}
&E_{\rm kin}=\int d\mathbf{x}\,\mathcal{E}_{\rm kin}(\mathbf{x})=\int d\mathbf{x}\frac{\hbar^2}{2m}\tau_0(\mathbf{x}),\label{kin}\\
&E_{\rm ext}=\sum_{t=0,1}\int d\mathbf{x}\, \rho_t(\mathbf{x})v_t(\mathbf{x}),\label{ext}
\end{align}
with $t = 0,1$ for the external fields in the isoscalar and isovector channels, respectively.
The external field in this work acts on the neutron and proton channels and takes the form in \cref{ExtField}.
In KS-DFT, the kinetic energy \cref{kin} is simply the sum of s.p. kinetic energies in the non-interacting system.

In this work, we consider the quasi-local Skyrme functional~\cite{Schunk2019}, i.e., the EDF only depends on local densities, including the number density $\rho(\mathbf{r})$, kinetic energy $\mathbf{\tau}(\mathbf{r})$, and spin-orbit density $\mathbf{J}(\mathbf{r})$, which are defined as \cite{Schunk2019}
\begin{align}
\rho(\mathbf{x}) =&\sum_{j}\left|\psi_{j}(\mathbf{x})\right|^{2},\\
\tau(\mathbf{x}) =&\sum_{j}\left|\nabla\psi_{j}(\mathbf{x})\right|^{2},\\
\mathbf J_z (\mathbf{x}) =&\sum_{j}\psi_{j}^{*}(\mathbf{x})\left(-i\right)\left(\nabla\times\boldsymbol\sigma\right)_{3}\psi_{j}(\mathbf{x}),
\end{align}
where $\psi_j$ is the KS orbitals and the label $j$ runs over all particles.
We follow the notations in Refs.~\cite{Marino2021, Marino2023}, in which the EDF takes the form
\begin{align}
\begin{aligned}
\mathcal{E}_{\rm pot}(\mathbf{x})=&\sum_{t=0,1}\Big[\sum_{\gamma}\left(c_{\gamma,0}+c_{\gamma,1}\beta^{2}\right)\rho_{0}^{\gamma+1}+C_t^\tau\rho_t\tau_t\\
&+C_t^{\Delta\rho}\rho_t\Delta\rho_t+C_t^J \mathbf{J}_t^2 + C_t^{\nabla J}\rho_t\nabla\cdot\mathbf{J}_t\Big]\label{PotDensity},
\end{aligned}
\end{align}
where $t = 0,1$ corresponds to the densities and parameters in the isoscalar and isovector channels, respectively.

Then, the KS orbitals of the ground state are obtained by solving the KS equation for the two kinds of particles ($q = n, p$ for neutrons and protons) \cite{Schunk2019}
\begin{align}
\begin{aligned}
\Big[&-\nabla\cdot\frac{\hbar^2}{2m_q^*(\mathbf{x})}\nabla+U_q(\mathbf{x})+v_q(\mathbf{x})\\
&+\mathbf{W}_q(\mathbf{x})\cdot(-i)\left(\nabla\times\boldsymbol{\sigma}\right)\Big]\psi_j(\mathbf{x})=\epsilon_j\psi_j(\mathbf{x}), \label{KSequation}
\end{aligned}
\end{align}
where
\begin{align}
U_q=\frac{\delta E}{\delta\rho_q},\quad\frac{\hbar^2}{2m_q^*}=\frac{\delta E}{\delta\tau_q},\quad\mathbf{W}_q=\frac{\delta E}{\delta\mathbf{J}_q}. \label{Def}
\end{align}
Here, $m_q^*(\mathbf{x})$, $U_q(\mathbf{x})$, and $\mathbf{W}_q(\mathbf{x})$ are the effective mass, KS potential, and spin-orbit potential, respectively. 
For a system with $A$ particles, the ground state is obtained from the $A$ lowest-energy solutions of Eq.~\eqref{KSequation}. 

Without loss of generality, in this work we consider external fields which are only functions of $z$. Thus, homogeneity is broken only along the $z$ direction, i.e.,
\begin{align}
\left\{ 
\begin{aligned}
&v_0(z)=2v_q\cos{(qz)},\\
&v_1(z)=2v_q\cos{(qz)}\hat{\tau}_z,\label{ExtField}
\end{aligned}
\right.
\end{align}
where $\tau_z$ is the isospin operator defined as 
\begin{align}
\tau_z = \begin{pmatrix} 1&0\\0&-1\end{pmatrix}.
\end{align}
With $\tau_z$, the external field $v_1$ imposes opposite potentials in the neutron and proton channels.
Hence, $v_1$ is called isovector external field.

Because of the PBC, also the quantity $q$ is an integer multiple of $\frac{2\pi}{L}$, i.e., $q = q_{\rm int} \frac{2\pi}{L}$ with integer $q_{\rm int}$.
The s.p. wavefunctions take the form:
\begin{align}
\label{eq: ks orbitals psi z}
\psi_{\mathbf n, q,\lambda}(\mathbf x)=\frac{e^{ik_xx}}{\sqrt{L}}\frac{e^{ik_yy}}{\sqrt{L}}\begin{pmatrix}\phi_{\mathbf n, q,\lambda}(z,\uparrow)\\\phi_{\mathbf n, q,\lambda}(z,\downarrow)\end{pmatrix}, 
\end{align}
where $k_x = \frac{2\pi}{L} n_x$ and $k_x = \frac{2\pi}{L} n_y$ are the momenta along the $x$- and $y$-axes.
Moreover, $\lambda = \pm 1$ is an additional quantum number analogous to the helicity and
represents the spin projection along the direction $n_x \hat{x} + n_y \hat{y}$ \cite{Marino2023, DANIELEWICZ200936}. 
As a consequence of \cref{eq: ks orbitals psi z}, the densities become functions of $z$ and are calculated in terms of the $\phi(z)$ orbitals as in Ref.~\cite{Marino2023}

Now, the KS equation can be recast into
\begin{align}
\begin{aligned}
&-\frac{d}{dz}\left(\frac{\hbar^{2}}{2m_q^{*}(z)}\phi_{\mathbf{n},q,\lambda}^{\prime}(z)\right)+\bigg(U_q(z)+v_q(z)\\
&+\lambda k_{n_xn_y}W_q(z)+\frac{\hbar^2}{2m_q^*(z)}k_{n_xn_y}^2\bigg)\phi_{\mathbf{n},q,\lambda}(z)=\epsilon_{\mathbf{n},\lambda}\phi_{\mathbf{n},q,\lambda}(z),\label{DFTequation}
\end{aligned}
\end{align}
where
\begin{align}
k_{n_xn_y} = \frac{2\pi}{L}(n_x \hat{x} + n_y \hat{y})\label{kxy}
\end{align}
is called transverse momentum \cite{Marino2023} and $\phi(z)$ is periodic.
The detailed derivation is given in Ref.~\cite{Marino2023}.

The equations for neutrons and protons are separated and solved independently in this work since the fields and orbitals are no longer identical for the two kinds of particles in the case of ANM.
Then, the explicit form for the effective mass, KS potential, and spin-orbit potential are calculated from the EDF in \cref{PotDensity}, for neutrons and protons, respectively (see App.~\ref{Details} for the details).

The KS equation is conveniently solved by expanding \cref{DFTequation} in the plane wave basis because of the intrinsic periodicity of the system, as in Refs.~\cite{Marino2023, e239da50c91a4a29bac0adb33d03568d}. 
The plane waves are constructed as 
\begin{align}
\phi_{\mathbf{n},q,\lambda}(z) = \frac{1}{\sqrt{L}}\sum_{k} c_k e^{ikz},
\end{align}
where, 
again, $k = 2n\pi/L$. 
Consequently, \cref{DFTequation} is recast into a matrix form:
\begin{align}
\label{eq: orbitals matrix form}
\sum_{k^{\prime}}\left(\tilde{h}_{\mathbf{n},\lambda}\right)_{k,k^{\prime}}c_{k^{\prime}}=\epsilon_{\mathbf{n},\lambda}c_{k},
\end{align}
where $\left(\tilde{h}_{\mathbf{n},\lambda}\right)_{k,k^{\prime}}$ is the Hamiltonian matrix in the plane-wave basis.

By solving the KS equation~(\ref{DFTequation}), the occupied orbitals and corresponding energies are obtained. 
The total energy is then evaluated in two different ways. The first one is by integrating the energy density \cref{PotDensity}:
\begin{align}
E=L^{2}\int_{-L/2}^{L/2}dz\,\mathcal{E}(z).\label{VolInt}
\end{align}
The second one is calculated by introducing the rearrangement energy \cite{Marino2021}
\begin{align}
E=\frac{1}{2}\left(T+\sum_{j}\epsilon_{j}\right)+E_{\rm rea},\label{TotalRea}
\end{align}
where 
\begin{align}
E_{\rm rea} = \int d\mathbf{r} \sum_{\gamma} \Big( \frac{1-\gamma}{2} \Big) (c_{\gamma,0} + \beta^2 c_{\gamma,1})\rho^{\gamma + 1}.
\end{align} 
The two ways of evaluating the total energy must match when the g.s.~solution is found.
It also serves as a check for the reliability and accuracy of the calculations.

The KS equation~(\ref{DFTequation}) is self-consistently solved, for neutrons and protons separately, for a given set of quantum numbers $(n_x,n_y,\lambda)$, and the quantum number $n_z$ labels the eigenstates with increasing eigenvalues.
The ground state is obtained by filling the lowest $N/g$ neutron states and $Z/g$ proton states. 
Moreover, the states with the same value of $n_x^2 + n_y^2$ are degenerate because the quantum numbers only enter the equation~(\ref{DFTequation}) by the transverse momentum (\ref{kxy}). 
Such degeneracy allows us to reduce the computational cost by restricting the solutions of equations to the orbitals with $0\leq n_x \leq n_y \leq n_{\rm max}$. 

\subsection{Response in asymmetric matter} \label{ReANM}
We summarize the basics of linear response theory at zero temperature, which is detailed, e.g.,  in~Refs.~\cite{PhysRevE.96.023203, Giuliani_Vignale_2005, Lundqvist1983TheoryOT}.
A static potential having the form of \cref{ExtField} is turned on in an ANM system described by Eq.~\eqref{DFTequation}.
The effect of the perturbation can be described at the leading order in the potential, as long as it is weak \cite{rin80}.
In a one-component system, the fluctuation of the density induced by the external potential is then given by
\begin{align}
\delta\rho(\mathbf{x})=\rho_{v}(\mathbf{x})-\rho_{0}=
\int d\mathbf{x}^{\prime}\chi(\mathbf{x},\mathbf{x}')v(\mathbf{x}'),\label{deltarho}
\end{align}
where $\chi(\mathbf{x},\mathbf{x}')$ is called the response function. 
In the present study, both the isoscalar and isovector responses are investigated. 
Therefore, the previous formula is generalized to the following matrix form:
\begin{align}
\begin{pmatrix} \delta\rho_0(\bf x) \\ \delta\rho_1(\bf x) \end{pmatrix} = 
\int d\mathbf{x}^{\prime}
\begin{pmatrix} 
\chi_{00}(\bf x,\bf x') & \chi_{01}(\bf x,\bf x') \\ 
\chi_{10}(\bf x,\bf x') & \chi_{11}(\bf x,\bf x')\end{pmatrix} 
\begin{pmatrix} v_0(\bf x')\\ v_1(\bf x')\end{pmatrix},\label{chimatrix}
\end{align}
where $\rho_0$ and $\rho_1$ are the isoscalar and isovector densities discussed in \cref{DFT} and $v_0$ and $v_1$ are the external fields in \cref{ExtField}. 
The response functions $\chi_{00}, \chi_{01}, \chi_{10}, \chi_{11}$ denote the response functions for the isoscalar and isovector densities with respect to the isoscalar and isovector fields, respectively.

To derive the relation between the energy and the strength of the fields, the energy functional~\eqref{totalenergy} is expanded with respect to the field around the unperturbed state $v_t = v_{t'} = 0$ as~\cite{nightingale1998quantum}
\begin{align}
\begin{aligned}
E[v]&-E[0]=
\sum_{t}
\int d\mathbf{x}\frac{\delta E}{\delta v_t(\mathbf{x})}\bigg|_{v_t=0}v_t(\mathbf{x}) \label{EvE0} \\
&+ \frac{1}{2} \sum_{t t^\prime} \int d\mathbf{x}\int d\mathbf{x}'\frac{\delta^{2}E}{\delta v_t(\mathbf{x})\delta v_{t^\prime}(\mathbf{x}')}\bigg|_{v_t= v_{t^\prime} = 0}v_{t}(\mathbf{x}) v_{t^\prime}(\mathbf{x}').
\end{aligned}
\end{align}
By recalling \cref{ext}, we notice that
\begin{align}
\frac{\delta E}{\delta v_t(\mathbf{x})} = \rho_t(\mathbf{x}). \label{deltaEv}
\end{align}
In particular, at the $v = 0$ limit, we find the uniform densities $\rho_t$.
Combining Eqs.~\eqref{chimatrix} and~\eqref{deltaEv}, the second variation of $E[v]$ reads
 \begin{align}
\frac{\delta^2E[v]}{\delta v_t(\mathbf{x})\delta v_{t^{\prime}}(\mathbf{x}')}=\frac{\delta\rho_t(\mathbf{x})}{\delta v_{t^\prime}(\mathbf{x}')}=\chi_{tt^\prime}(\mathbf{x},\mathbf{x}').
\end{align}
Therefore, \cref{EvE0} can be written as
\begin{align}
E[v]-E[0]=& \sum_{tt^\prime}\int d\mathbf{x}\,v_t(\mathbf{x})\rho_{t}\\
&+{\frac{1}{2}}\int d\mathbf{x}\int d\mathbf{x}^{\prime} v_t(\mathbf{x}) \chi_{tt^\prime}(\mathbf{x},\mathbf{x}^{\prime}) v_{t^\prime}(\mathbf{x}^{\prime}).\label{enresponse}
\end{align}

Since any general periodic function can be decomposed into a superposition of plane waves, we consider a monochromatic potential oscillating with a given transferred momentum $\mathbf{q}$:
\begin{align}
v(\mathbf{x})=v_{q}e^{i\mathbf{q}\cdot\mathbf{x}}+c.c.=2v_{q}\cos\left(\mathbf{q}\cdot\mathbf{x}\right).\label{Extfield}
\end{align}
Then, the matrix \cref{chimatrix} for the response functions in ANM is recast into
\begin{align}
\begin{pmatrix} \rho_{q,0} \\ \rho_{q,1} \end{pmatrix} = \begin{pmatrix} \chi_{00}(q) & \chi_{01}(q) \\ \chi_{10}(q) &\chi_{11}(q)\end{pmatrix} \begin{pmatrix} v_{q,0}\\ v_{q,1}\end{pmatrix},\label{chimatrix1}
\end{align}
with a Fourier transformation .

Similarly, the Fourier transformation is also performed to calculate the energy response and we have
\begin{align}
\delta e_{v}=e_{v}-e_{0}= \sum_{t}\frac{\chi_{tt}(q)}{\rho_{q,t}}v_{q,t}^{2},
\label{enr}
\end{align}
where $e_v = E[v]/A$ is the energy per particle.

Therefore, response functions can be extracted in two ways, i.e., either from the variation of the densities or from the variations of the energy. Such variations are evaluated by changing the strength and the momentum of the perturbation, given a specific isospin channel of this latter.

To benchmark the DFT methods in \cref{DFT}, the response functions $\chi(q)$ obtained from the numerical finite-$A$ calculations will be compared with the response functions calculated with linear response theory in the TL.
In particular, response functions in the random phase approximation (RPA) are discussed.
By summing the series of bubble diagrams (see Refs.~\cite{rin80,fetter2003quantum}) and describing the system's response in terms of excitations made by superposing particle-hole pairs, as a result of the residual \textit{ph} interaction, 
RPA is exact as long as the external field is weak, i.e., RPA is an exact implementation of linear response theory.
Therefore, the FS effects can be tested by comparing the results 
in the finite-$A$ system to those obtained using RPA.
In fact, RPA response functions in the TL are known analytically (see App.~\ref{RPAresponse}) and can be used as a benchmark for the DFT method of Sec.~\ref{DFT} and to test the sensitivity of the response functions with respect to the EDF parameters. 
The method to calculate the response functions with RPA is given in App.~\ref{RPAresponse}.
Large-$A$ calculations with our DFT method should converge to the RPA results, as long as the external fields are small.

\section{Validation of the method}\label{NT}
\begin{figure}[t]
\centering
\includegraphics[width = 9cm]{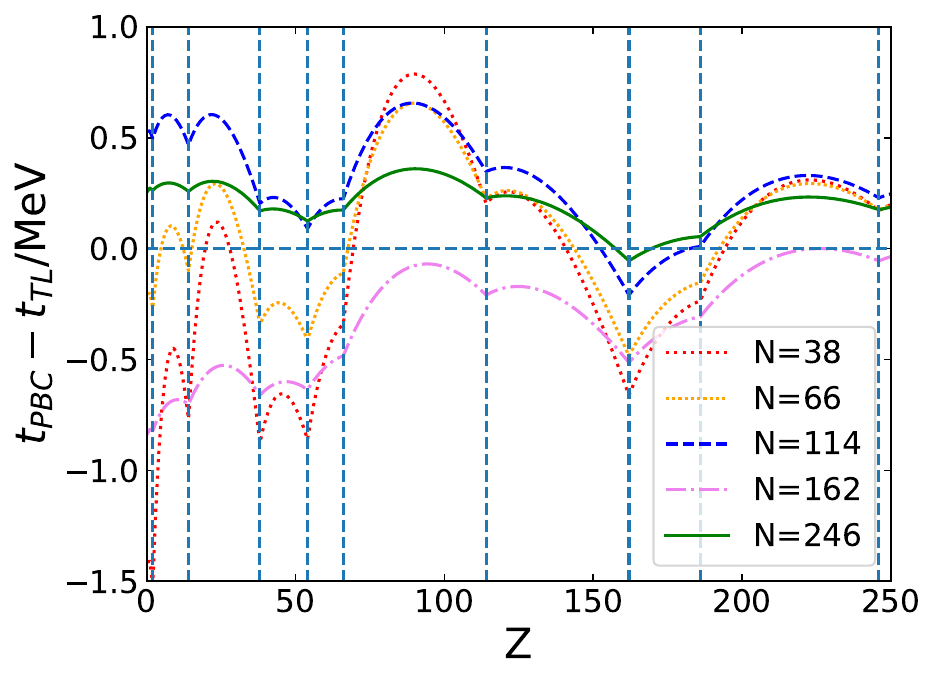}
\caption{Difference between the kinetic energy per particle of the Fermi gas with a finite number of particles and that in the TL as a function of the number of protons in the finite-$A$ system.
Different curves show the results for different numbers of neutrons.
The vertical dashed lines denote the ``magic numbers'' in the momentum space.}
\label{kin_NZ}
\end{figure}

In this section, we perform a validation of our method by investigating the EoS and the static response of ANM, with a particular emphasis on discussing FS effects.
To ensure the translational invariance, the number of particles in the box should correspond to one of the momentum-space ``magic numbers'', namely to a closed-shell configuration.
In SNM or PNM, multiples of 33 particles are commonly used because the kinetic energy of the free Fermi gas with $33g$ particles is close to that of the TL gas \cite{PhysRevC.95.044309}, which means the FS effects should be small.
In the present study, the calculations are extended to the ANM.
Thus, we first discuss the choice of the number of particles in ANM.
The kinetic energy per particle $t$ is calculated in both a finite-$A$ gas and in the TL.
In TL, the kinetic energy is calculated as~\cite{Marino2021}
\begin{align}
\begin{aligned}
&t(\rho,\beta) = \frac{t_{\rm sat}}{2}[(1+\beta)^{\frac{5}{3}} + (1-\beta)^{\frac{5}{3}}] \left(\frac{\rho}{\rho_{\rm sat}}\right)^{\frac{2}{3}},\\
&t_{\rm sat} = \frac{3\hbar^2}{10m}\left(\frac{3\pi^2}{2}\right)^{\frac{2}{3}} \rho_{\rm sat}^{\frac{2}{3}},
\end{aligned}
\label{t}
\end{align}
where $\rho_{\rm sat} = 0.16\,\rm{fm}^{-3}$ is the nuclear saturation density.
The deviation $t_{\rm PBC}-t_{\rm TL}$ for the free Fermi gas is shown in \cref{kin_NZ} as a function of the number of protons $Z$, with different curves referring to different "magic numbers" of neutrons $N$.
Vertical dashed lines refer to proton shell closures.

To our knowledge, the finite-$A$ and TL kinetic energies in asymmetric matter were not compared previously.
Overall, the difference between the two kinetic energies tends to decrease as the number of particles increases.
For a given $N$, however, the discrepancy fluctuates significantly as a function of the proton number in between two ``magic numbers'', being largest when the outermost proton shell is approximately half-filled and much smaller in correspondence of a proton shell closure. 
In particular, we have noticed that $N = 186$ and  $Z = 114$ is a good choice, as it yields a kinetic energy which is very close to that of the TL, while still being manageable computationally within our DFT calculations.

\begin{figure}[t]
\centering
\includegraphics[width = 8cm]{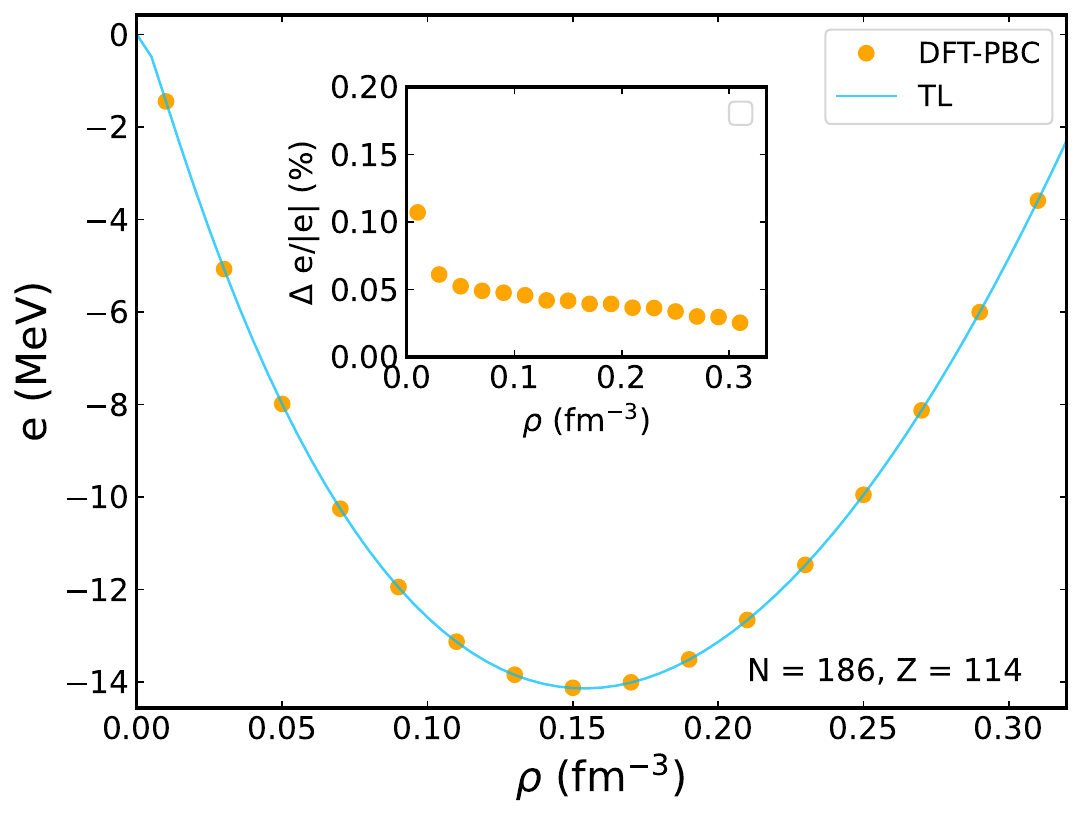}
\caption{EoS for ANM with $\beta = 0.24$ calculated with the SLy4 EDF.
The results calculated within PBC with $N = 186$ and $Z = 114$ are shown with the dots. 
For comparison, the corresponding results of the thermodynamic limit (TL) with the same $\beta$ are shown with the solid line. 
Inset: The relative difference $|e_{\rm PBC}-e_{\rm TL}|/|e_{\rm TL}|$ between the results of the finite system and TL.}
\label{EOS300}
\end{figure}

This preliminary analysis of the free gas kinetic energy is now followed by a validation of our approach for interacting matter.
First, in \cref{EOS300}, the EoS is reported as a function of the total density for the case of $N = 186$ and $Z = 114$, corresponding to an isospin asymmetry of $\beta = 0.24$, and compared to predictions for the TL EoS at the same $\beta$, which is given by
\begin{align}
e(\rho,\beta) = t(\rho, \beta) + v(\rho,\beta),\label{eosequation}
\end{align}
where the kinetic energy $t$ is calculated with \cref{t} and 
\begin{align}
v(\rho,\beta) = \sum_{\gamma}c_\gamma(\beta)\rho^\gamma = \sum_{\gamma}[c_{\gamma,0}+c_{\gamma,1}\beta^2]\rho^\gamma.
\end{align}
Here, $\gamma$ refers to the powers entering the potential part of the EDF, see Eq.~(\ref{PotDensity}).
It is found that the numerical finite-$A$ results and the TL EoS in \cref{eosequation} agree well at every point, as can be seen in the inset, which shows that the relative error drops well below 0.1\% already for densities above 0.05 fm$^{-3}$.
This illustrates that the choice of $N=186$ and $Z=114$ efficiently reduces the FS effects also for full DFT computations.

\begin{figure}[t]
\centering
\includegraphics[width = 8.5cm]{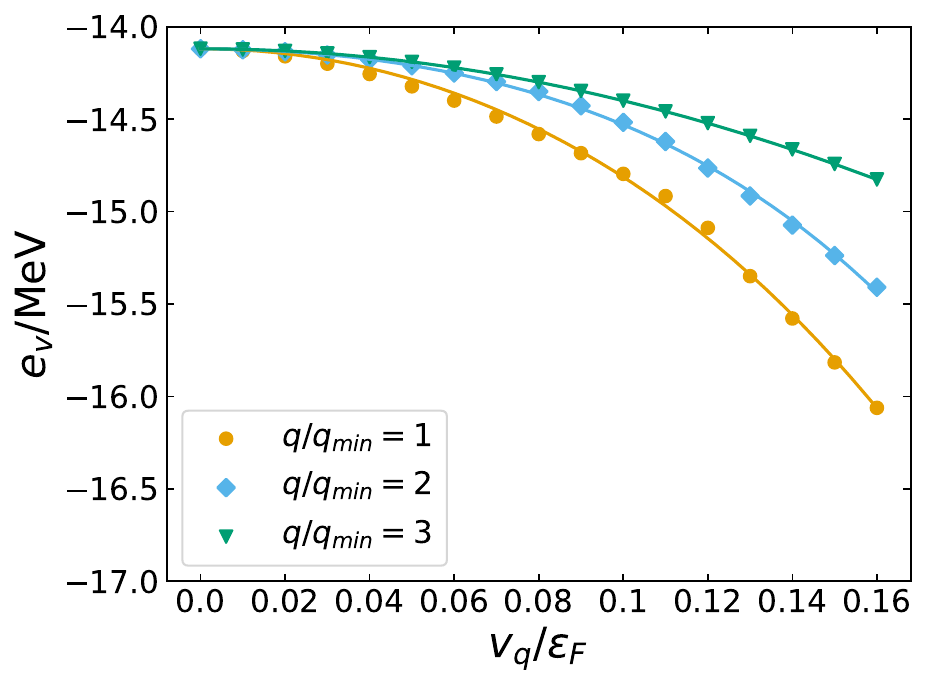}
\caption{Energy per particle as a function of the strength of the external fields in the isoscalar channel, calculated in systems with $N= 186$ and $Z = 114$.
The transferred momenta of the external fields are $q/q_{\rm min} = 1,2,3$.
The curves show the fitting results for the numerical data.
}
\label{EnergyFit}
\end{figure}

Next, we discuss our predictions for ANM subject to an external perturbation.
The energies of the perturbed system can be studied as a function of the strength and the momentum of the external field:
\begin{align}
e_v = e_0 + \frac{\chi(q)}{\rho_{q,t}} v_{q,t}^2 + c_4 v_{q,t}^4, \label{EnResponse}
\end{align}
where both $c_4$ and $\chi(q)/\rho_{q,t}$ are coefficients to be fitted.
We use \cref{EnResponse} rather than the quadratic relation \cref{enr} discussed in \cref{ReANM}, because the fourth-order term is found to be non-negligible in numerical calculations as the external fields have finite magnitudes.

The response functions $\chi(q)/\rho$ can be extracted from the behavior of the energy per particle at fixed momentum $q$.
For each $q$ value, the calculations are carried out by different strengths $v$ ranging from $0.01\epsilon_F^N$ to $0.16\epsilon_F^N$, where $\epsilon_F^N$ is the Fermi energy of neutrons.
The fitting is then performed by the least squares method, and the results for the ratios $-\chi (q) / \rho_0$ and their errors are obtained.

In \cref{EnergyFit}, the results of the calculations and the interpolating functions are reported in the case of $N = 186$ and $Z = 114$, for an isoscalar perturbation [$t=0$ in Eq.~\eqref{EnResponse}].
The quadratic fit is accurate for $q > 2q_{min}$, while for $q \leq 2q_{\rm min}$ some small deviations are observed, especially when the strength of the external fields is large.
Also, as we shall see, the isoscalar response has larger magnitude when $q$ is smaller, suggesting that the perturbative regime is valid in a more limited range of strengths.

\begin{figure}[t]
\includegraphics[width = 8.5cm]{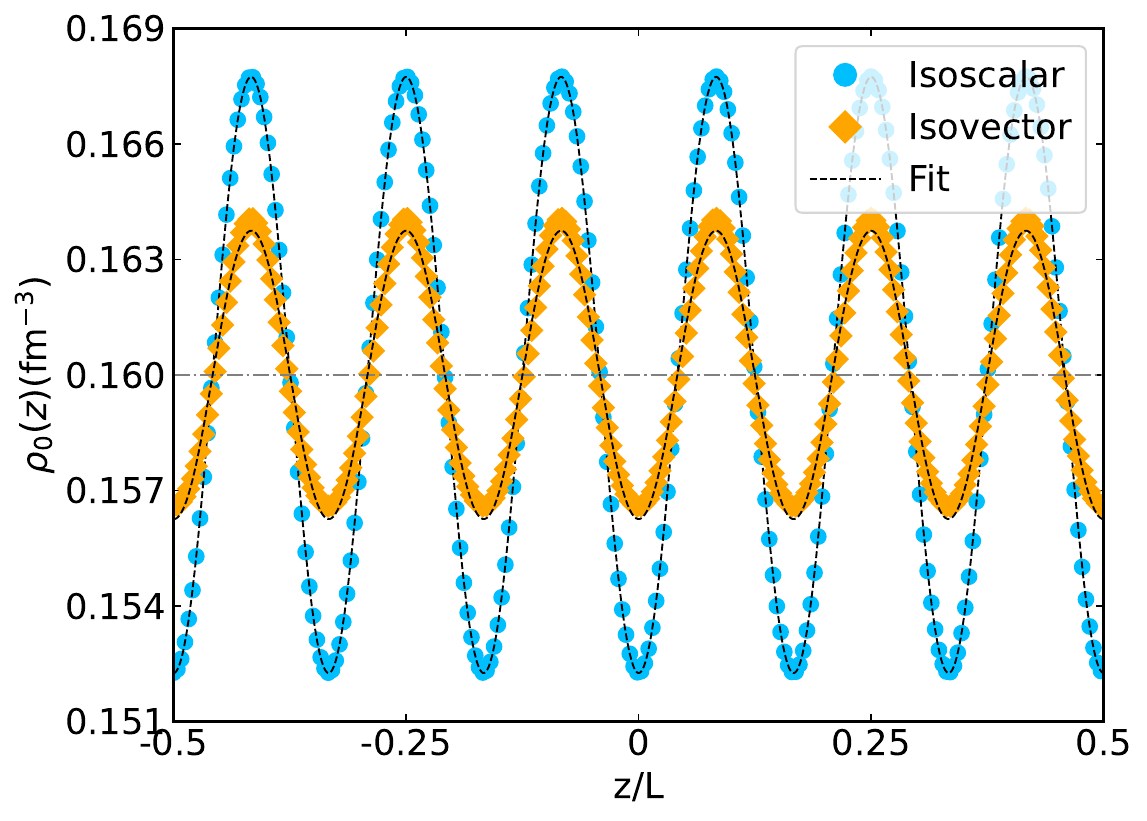}
\caption{Perturbed isoscalar densities of a system with $N=186$ and $Z=114$ in the presence of external fields with the transferred momentum $q = 6q_{\rm min}$ and strength $v_q = 0.15 \epsilon_F$. 
Results for the case when an isoscalar (isovector) external field is applied are shown as blue dots (orange diamonds).
The data are fitted to cosine functions, shown as black dashed lines.
The grey dashed line refers to the reference isoscalar density of the unperturbed system.}
\label{PD300IS}
\end{figure}

\begin{figure}[t]
\includegraphics[width = 8.5cm]{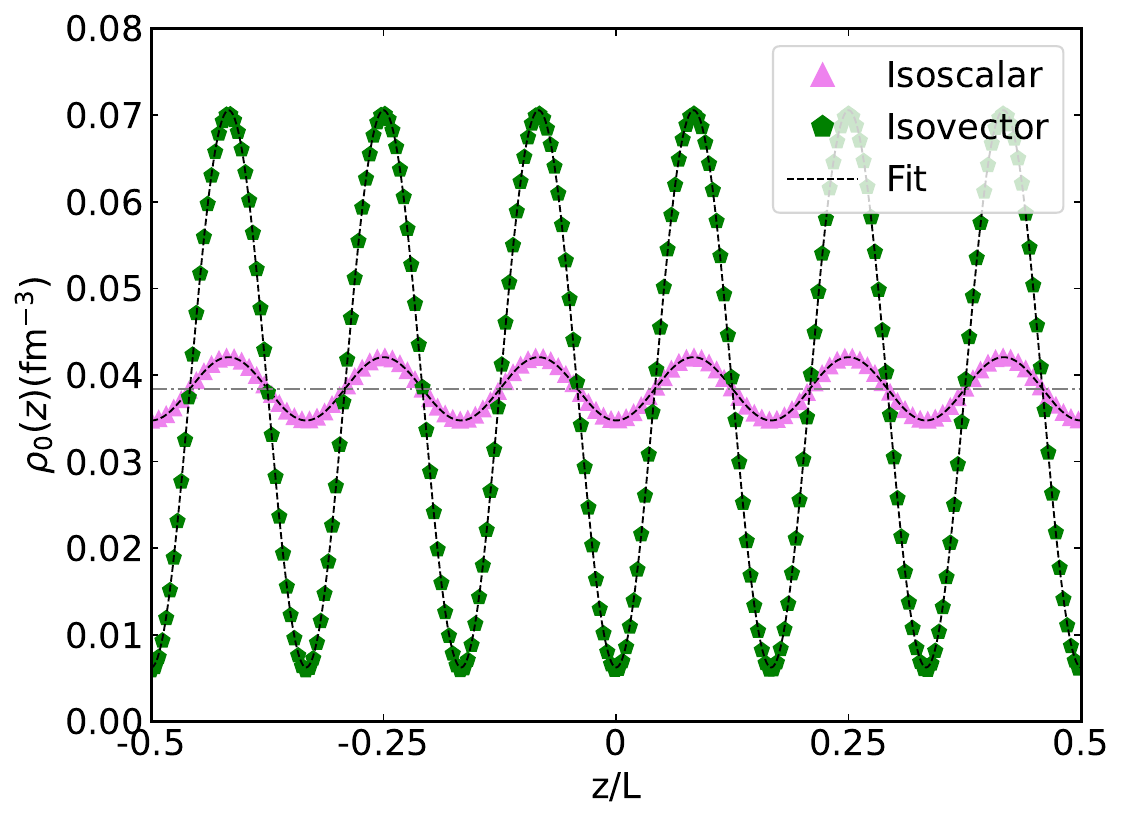}
\caption{Same as \cref{PD300IS}, but for the perturbed isovector density.
Here, the grey line shows the initial isovector density.
Because the isoscalar external field imposes opposite potentials in the neutron and proton channels, the fluctuation of the isovector density under the perturbation of an isovector external field turns out to be very large.}
\label{PD300IV}
\end{figure}

\begin{table}[b]
\begin{center}
\caption{Fourier components of the perturbed isoscalar densities of the systems in the presence of the external fields with different periodicity $q/q_{min}$ and same strength in both the isoscalar and isovector channels.
The type of the external fields is indicated with "IS" and "IV" for the isoscalar and isovector channels, respectively.
The results are shown in logarithmic scale, $\log_{10}(A_n/A_{n_{\rm dom}})$.\label{Table0}
}
\begin{tabular}{cccccccccccc}
 \toprule
  \multirow{2}{*}{$q/q_{\rm min}$}& \multirow{2}{*}{Type}&\multicolumn{3}{c}{$\log_{10}(A_n/A_{n_{\rm dom}})$}\\
  \cmidrule{3-5}
   & &$n=2 n_{\rm dom}$&$n=4 n_{\rm dom}$&$n=6 n_{\rm dom}$\\
   \midrule
    2&IS&$-2.59$&$-6.46$&$-9.92$\\
    4&IS&$-8.04$&$-11.49$&$-13.32$\\
    6&IS&$-6.86$&$-10.61$&$-17.11$\\
    2&IV&$-3.64$&$-4.99$&$-7.14$\\
    4&IV&$-1.50$&$-4.81$&$-6.54$\\
    6&IV&$-2.50$&$-5.81$&$-10.96$\\

\bottomrule
\end{tabular}
\end{center}
\end{table}

\begin{table}[b]
\caption{Same as \cref{Table0}, but for the results for the perturbed isovector densities.\label{Table1}}
\begin{center}
\begin{tabular}{cccccccccccc}
 \toprule
  \multirow{2}{*}{$q/q_{\rm min}$}& \multirow{2}{*}{Type}&\multicolumn{3}{c}{$\log_{10}(A_n/A_{n_{\rm dom}})$}\\
  \cmidrule{3-5}
   & &$n=2 n_{\rm dom}$&$n=4 n_{\rm dom}$&$n=6 n_{\rm dom}$\\
   \midrule
   	2&IS&$-1.14$&$-3.82$&$-3.93$\\
    4&IS&$-4.56$&$-9.57$&$-14.96$\\
    6&IS&$-5.16$&$-11.69$&$-16.80$\\
    2&IV&$-3.36$&$-4.36$&$-8.92$\\
    4&IV&$-3.81$&$-5.41$&$-8.20$\\
    6&IV&$-4.47$&$-7.80$&$-12.74$\\

\bottomrule
\end{tabular}
\end{center}
\end{table}

We further discuss the response by considering the isoscalar and isovector densities of the perturbed system (for the case of $N = 186$ and $Z = 114$), in the presence of external fields, in Figs.~\ref{PD300IS} and \ref{PD300IV}, respectively.
We consider the case where a purely isoscalar (isovector) perturbation is applied to the system and show the resulting density with dots (diamonds).
As a consequence of PBC, the transferred momenta of the external potentials is an integer multiple of the smallest allowed transferred momentum, i.e., $q = nq_{\rm min} = n \frac{2\pi}{L}$, where $n$ is an integer and $L$ is the size of the box.
In the plots, we have considered the case of $q=6q_{\rm min}$ and strength  $v_q = 0.15 \epsilon_F^N$. 

Due to the perturbation, both the isoscalar and isovector density turn out to be periodic functions of $z$.
Fits to cosine functions with the same wavelength as the external fields are shown as black dashed lines.
It can be noticed that the cosine interpolation reproduces the data accurately, suggesting that, in the case considered, linear response theory works well.
In fact, the contribution of higher-order harmonics is small, and densities are dominated by a single Fourier component.

The amplitudes of the density fluctuation in the isoscalar and isovector channels $\rho_{q,0}$ and $\rho_{q,1}$, see Eq.~\eqref{chimatrix1}, are then discussed.
In Fig.~\ref{PD300IS}, where $\rho_{0}(z)$ is shown, the fluctuation is larger when the isoscalar external field is applied ($\rho_{q,0} \approx 0.008 \rm fm^{-3}$), because, on an intuitive level, the perturbation is acting on the two kinds of nucleons in phase, while the isovector external fields have opposite effects on protons and neutrons.
In a specular way, an isovector external field (Fig.~\ref{PD300IV}) induces a large fluctuation in the isovector density, with $\rho_{q,1} \approx 0.03 \rm\,fm^{-3}$, while its effect on the isoscalar density is small.

The validity of the assumption of linear response is discussed in more detail by decomposing the perturbed density $\rho_t$ into a Fourier series,
\begin{align}
\rho_t = \sum_{n} A_n \cos (\frac{2\pi}{L}nz). \label{FourierComponents}
\end{align}
In Tables~\ref{Table0} and \ref{Table1}, we report the contributions of different Fourier components $A_n$ to the isoscalar and isovector densities, respectively, for the case of different periodicities $q/q_{\rm min}$ and same strength $v_q = 0.15\epsilon_F^N$ of the external response (either of iscoscalar or isovector type).
For convenience, we normalize the Fourier coefficients to the expected dominant component $A_{n_{\rm dom}}$, corresponding to the periodicity of the external field, i.e., $n_{\rm dom} = q/q_{\rm min}$, and list them in logarithmic scale, namely we show $\log_{10}(A_n/A_{n_{\rm dom}})$.
We only show $A_{n}$ with $n$ equals to even multiples of $n_{\rm dom}$ because the Fourier components with odd multiples of $n_{\rm dom}$ are found to be negligible.

In general, the Fourier components with the same period as the external fields are dominant.
When the transferred momentum is small, e.g., $q/q_{\rm min}=2$, some non-dominant components are not negligible, which explains the deviations in \cref{EnergyFit}.
As the transferred momentum increases, the dominant component becomes much larger than the others.
We conclude that the quantity $\rho_q$ can be correctly extracted by fitting the densities to cosine functions with the same periodicity as the external potential.
Response functions $\chi(q)$, then, can be obtained accurately by studying $\rho_q$ for small perturbation strengths, as long as $q/q_{\rm min} > 2$. (For smaller momenta, some contamination from higher-order components may appear.)

\section{Results}\label{results}

\begin{figure}[t]
\centering
\includegraphics[width = 8.5cm]{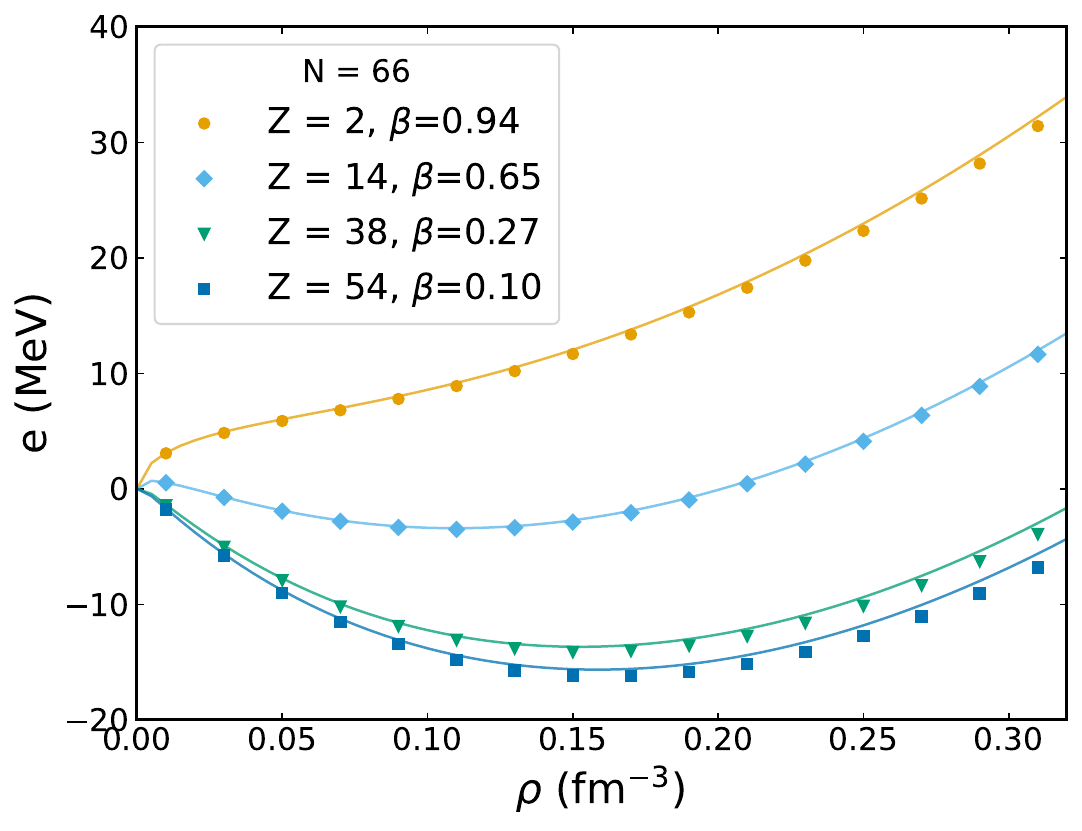}
\caption{EoS for systems with $N = 66$ and different numbers of protons calculated with the SLy4 EDF.
The corresponding isospin asymmetry is also reported in the legend.
}
\label{EoS}
\end{figure}

The DFT method discussed in \cref{DFT} is applied to calculate the EoS, energy levels, and response functions of ANM.  
Numerical calculations are performed with PBCs using the SLy4 EDF \cite{CHABANAT1997710}.
To study systems with different isospin asymmetries $\beta$, we fix the number of neutrons to $N = 66$ and vary the number of protons.
If not stated otherwise, the reference density is set to $\rho_0 = 0.16~\rm{fm}^{-3}$.
The strengths of the external fields are measured in units of the Fermi energy of neutrons $(v_q/\epsilon_F^N)$.
The transferred momenta of the external perturbations are multiples of the minimum allowed momentum, $q=q_{\rm int}2\pi/L$ with $q_{\rm int}$ being an integer number.

\subsection{Equation of state}
\label{secEOS}

In \cref{EoS}, we report the ANM EoS at different isospin asymmetries.
TL calculations are shown as solid lines.
Finite-$A$ results are obtained keeping the number of neutrons fixed to 66 and varying $Z$ with different magic numbers.
Overall, the two sets of computations agree reasonably well.
Discrepancies tend to increase slightly with increasing $\rho_0$.
Also, they are larger when $Z = 38$ and $54$, which is consistent with the calculations of kinetic energy per particle in \cref{kin_NZ}.
In general, the free gas picture offers a useful guide to identify the particle numbers that allow to reduce the FS effects in interacting ANM too.

\subsection{Energy Levels}
\label{sec: energy levels}

\begin{figure*}[t]
\centering
\includegraphics[width = \textwidth]{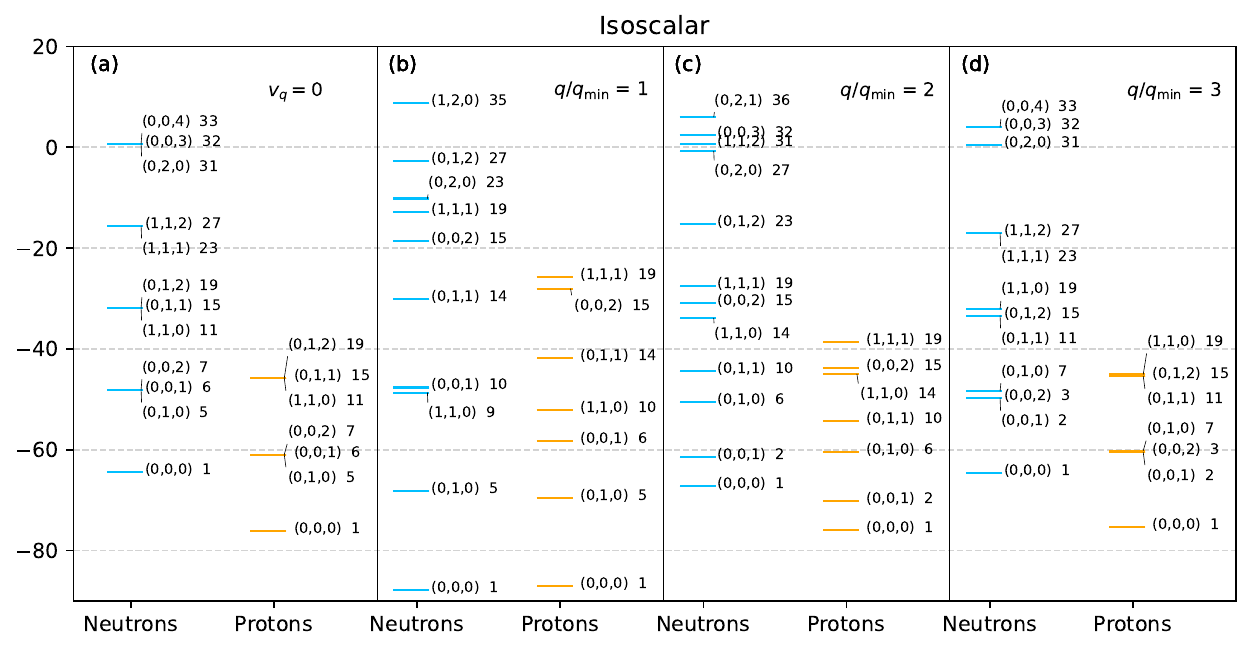}
\caption{Energy levels for ANM with 66 neutrons and 38 protons subject to an isoscalar external perturbation.
The energy levels for neutrons and protons are shown in blue and orange, respectively.
We focus on the spin-up case $\lambda = 1$.
The quantum numbers $(n_x,n_y,n_z)$ and the corresponding number of particles up to closed shells are shown beside each level.
(a) Energy levels of the unperturbed system.
(b)-(d) Energy levels of the systems in the presence of three isoscalar external fields with the same strength $v = 0.25 \epsilon_F$ and increasing transferred momentum $q/q_{\rm min}$.
}
\label{EL0}
\end{figure*}

\begin{figure*}[t]
\centering
\includegraphics[width = \textwidth]{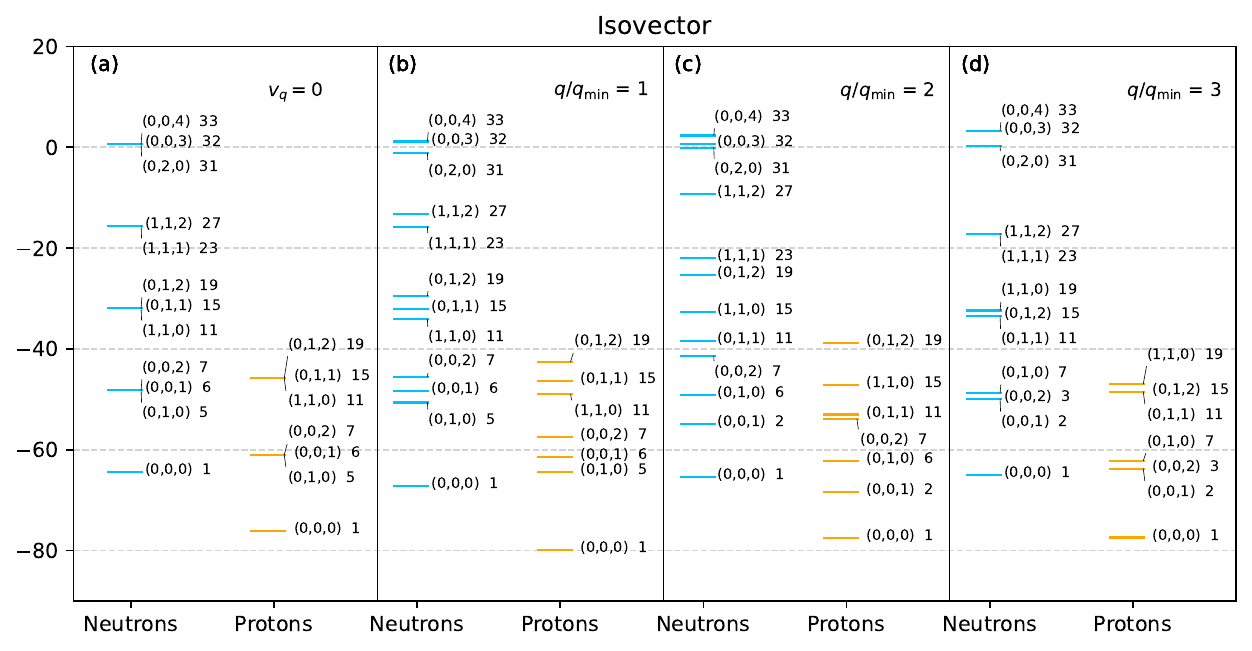}
\caption{Same as \cref{EL0}, but for the case in which an isovector perturbation is applied.
}
\label{EL1}
\end{figure*}

Next, we investigate the evolution of the s.p. level structure in the presence of external perturbations.
Momentum-space magic numbers are known for uniform systems.
As pointed out in Ref.~\cite{Marino2023}, however, the shell structure may change when relatively strong external potentials are applied.
In the present work, we study the shell structure of perturbed ANM.
As a representative case, we focus on a system with 66 neutrons and 38 protons subject to PBC.
In Fig.~\ref{EL0} (\ref{EL1}), we apply an isoscalar (isovector) field with strength $v_q = 0.25\epsilon_F^N$, and show the proton and neutron s.p. levels for different values of the transferred momentum $q/q_{\rm min}$ in panels (b)-(d).
For comparison, the energy levels in the unperturbed case ($v_q=0$) are reported in panel (a).
Since states with $\lambda = \pm 1$ are degenerate, we only focus on $\lambda = 1$.
Beside each level, we show its quantum numbers $(n_x,n_y,n_z)$, with $n_x \leq n_y$, and the number of particles corresponding to shell closures.
In the perturbed system, in general, the degeneracy of a given state is related only to the invariance of s.p. energies under the exchange of $n_x$ and $n_y$ or of the sign of $n_x$ or $n_y$.
For $n_x=n_y=0$, the multiplicity is 1.
States with $n_x = 0$ but $n_y \neq 0$ have multiplicity 2, while, when both $n_x$ and $n_y$ are different from zero, the degeneracy is equal to 4.
Further symmetries characterize the unperturbed system, as discussed in Sec.~\ref{ANM}, and show up in additional degeneracies of the energy levels.

We first discuss the case of an isoscalar perturbation in Fig.~\ref{EL0}.
When the transferred momentum is small, i.e., $q/q_{\rm min} = 1$ and $2$, the degeneracies of the unperturbed system are lifted, and we observe a significant rearrangement of the level ordering for both neutrons and protons.
In particular, the neutron magic number $27$ disappears in \cref{EL0}(b) and in \cref{EL0}(c) is replaced by the new numbers $23$.
When $q/q_{\rm min} = 3$, see \cref{EL0}(d), the degeneracy is lifted only partially.
For example, the triplet of states around 0 MeV in the homogeneous case, panel (a), splits into the $(0,2,0)$ level and a doublet of degenerate states, $(0,0,3)$ and $(0,0,4)$, while a shell closure at $N=27$ persists.
Proton levels up to $Z=19$ are overall less impacted by the perturbation. While the closure for 7 particles might vanish, 19 remains a magic number, and, at $q/q_{\rm min} = 3$, levels show the same degeneracies as in the unperturbed system.

Then, in Fig.~\ref{EL1} we consider the case of an isovector perturbation.
Degeneracies are also lifted when $q/q_{\rm min} = 1$ and $2$. However, the ordering of the neutron energy levels is overall kept, and a neutron magic number persists at $N = 27$.
Also, proton energy levels are shifted by less than $10$ MeV wrt.~the unperturbed system, indicating that the isovector perturbation has a lesser impact than the isoscalar one.

To sum up, our study suggests that the impact of external perturbations on the shell structures is larger in the isoscalar channel and for small values of the momentum $q/q_{\rm min}$.
This behavior is further investigated in App.~\ref{Energychange}.

\subsection{Finite-size effects}

In this section, we investigate the ANM response functions and the magnitude of the FS effects as a function of the isospin asymmetry and the transferred momentum.
First, we exploit the matrix equation~\eqref{chimatrix1} to determine the four response functions $\chi_{00},\chi_{01},\chi_{10},\chi_{11}$. 
This formula provides a straightforward linear relation between the perturbed densities, perturbation strength, and the response functions. 
All the four response functions can be extracted by considering the effect of isoscalar/isovector perturbations with strength $v_{q,0}$/$v_{q,1}$, respectively on isoscalar/isovector densitities $\rho_{q,0}$ and $\rho_{q,1}$.
In practice, we include a perturbation with given momentum $q$ and strength $v_q$ and determine numerically the density 
Then, we extract the response function $\chi_{tt^{\prime}}$ by evaluating the ratio $\rho_{q,t}/v_{q,t^{\prime}}$, 
which converges to $\chi_{tt^{\prime}}$ in the limit of vanishing strength, for finite but small values of the strength.
We have performed calculations with $N = 66$ and different proton numbers, keeping the total density at $\rho_0 = 0.16 \rm fm^{-3}$ and considering strengths $v_q = 0.08\sim 1.60 \epsilon_F$ and transferred momentum $q_{\rm int} = 1 \sim 20$.
FS effects are quantified by considering the deviations between finite-$A$ and RPA results.

\begin{figure}[h]
\centering
\includegraphics[width =\columnwidth]{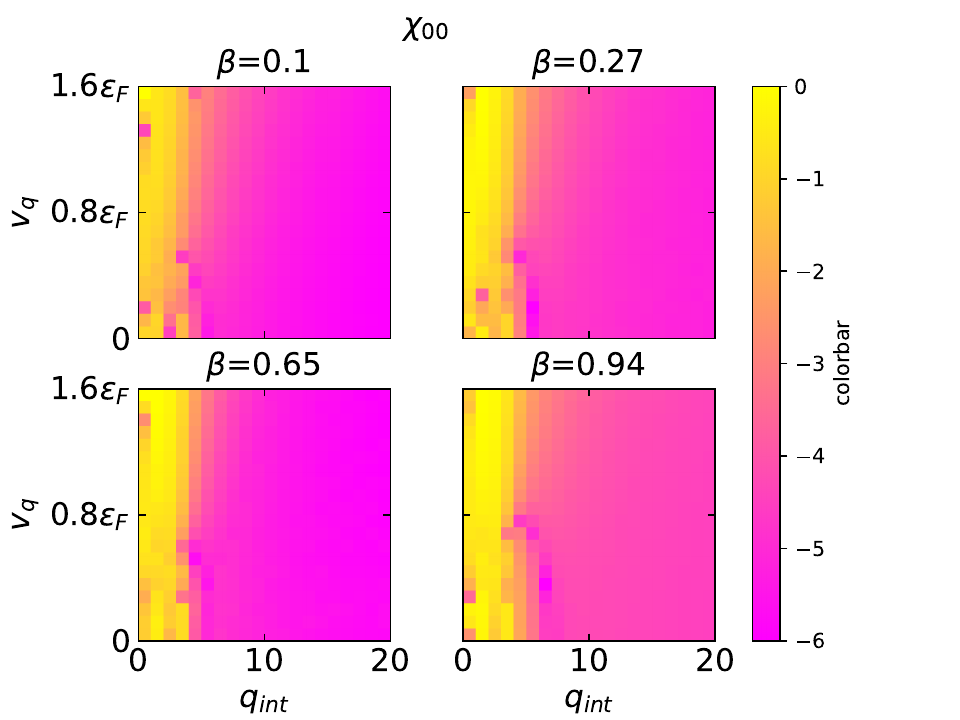}
\caption{Relative deviation $|\chi_{\rm PBC} - \chi_{\rm PRA}|/\chi_{\rm RPA}$ (in logarithmic scale) between the RPA and finite-$A$ (denoted as "PBC") predictions for the response function $\chi_{00}$, represented as a two-dimensional color map (see the colorbar on the right for the color scale).
The strength and transferred momentum of the external fields are shown on the vertical and horizontal axes, respectively.
Calculations are performed at density $\rho_0 = 0.16\  \rm fm^{-3}$, while each panel refers to a different isospin asymmetry.
PBC results are obtained with 66 neutrons while varying the proton number. See the text for details.
}
\label{chi00}
\end{figure}

\begin{figure}[b!]
    \centering
    \includegraphics[width=\columnwidth]{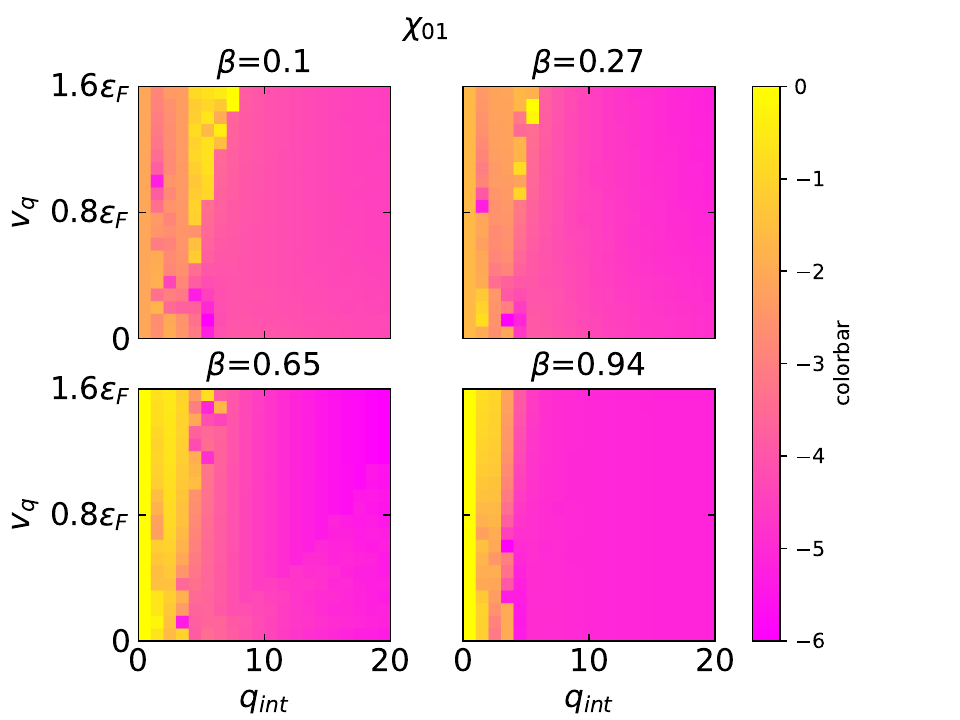}
    \caption{Same as \cref{chi00}, but for the response function $\chi_{01}$.}
    \label{chi01}
\end{figure}

\begin{figure}[h!]
\centering
\includegraphics[width =\columnwidth]{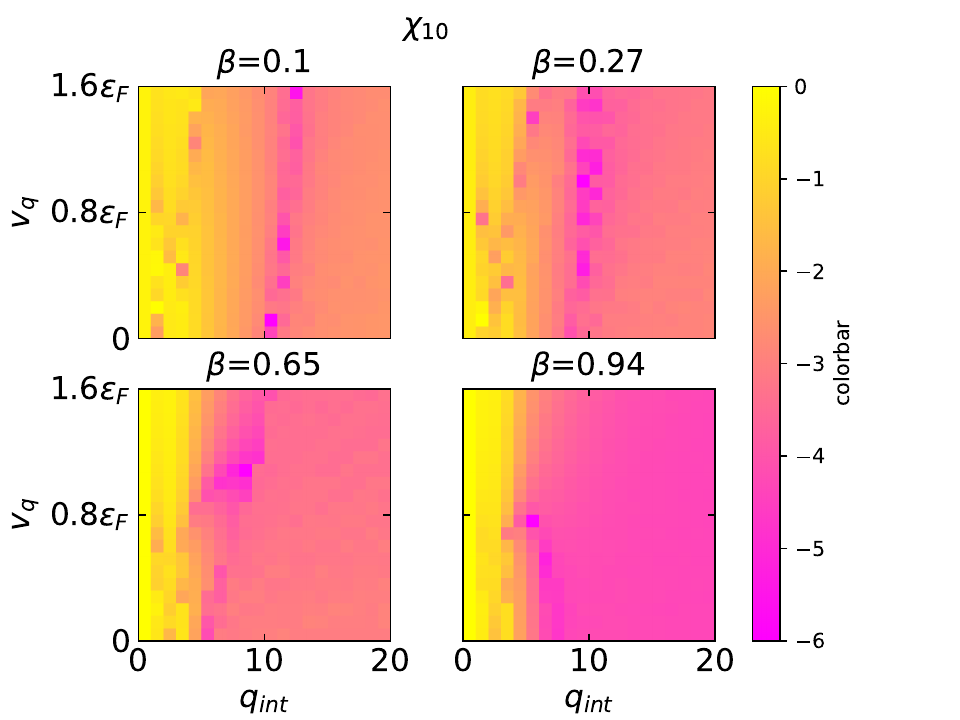}
\caption{Same as \cref{chi00}, but for the response function $\chi_{10}$.}
\label{chi10}
\end{figure}

\begin{figure}[b!]
    \centering
    \includegraphics[width=\columnwidth]{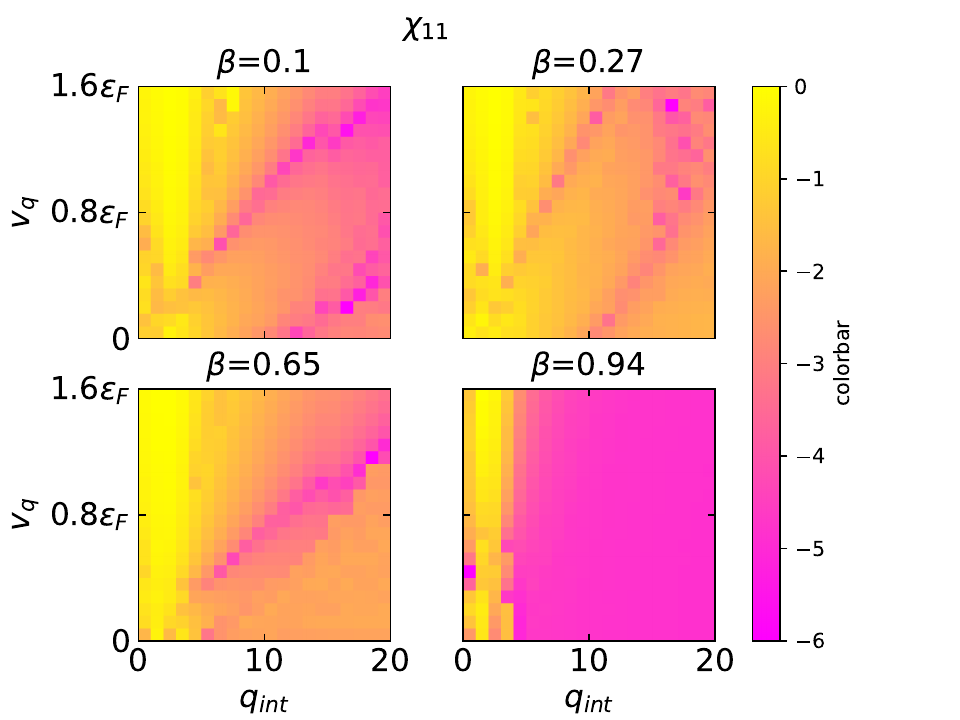}
    \caption{Same as \cref{chi00}, but for the response function $\chi_{11}$.}
    \label{chi11}
\end{figure}

In \cref{chi00,chi01,chi10,chi11} we display the deviations between the two results for the different response functions, with each panel referring to a different $Z$ (and thus a different isospin asymmetry).
The two-dimensional maps represent the relative difference between calculations with RPA and PBC, $\log_{10}(|\chi_{\rm PBC} - \chi_{\rm PRA}|/\chi_{\rm RPA})$, as a function of the strength and transferred momentum of the external field.
A color bar is shown to the right side of each figure.
$\chi_{\rm PRA}$ is evaluated analytically at momentum $q = 2\pi/L q_{\rm int}$ (see App.~\ref{ReANM}) and is independent of $v_q$.
$\chi_{\rm PBC}$ refers to the ratio between the density amplitude and the perturbation strength for the proper combination of isospin indices.
Strictly speaking, the response function is found only in the limit of vanishing strength.
In practice, we expect the ratio to be approximately constant in a finite region of strengths.

For the response function $\chi_{00}$ (\cref{chi00}), the finite-$A$ and TL results overall agree well when $q_{\rm int}>4$ for all values of $\beta$.
FS effects tend to increase with larger $v_q$, e.g. for $5\leq q_{\rm int}\leq 8$, indicating that the linear relation between the density amplitudes and perturbation strengths is breaking down, especially when $v_q > 0.8\epsilon_F$.
Also, FS effects are apparently larger and inconsistent with different strengths $v_q$ at small momenta $q_{\rm int}$.
We have already noticed that, in general, the perturbations may cause a rearrangement of the energy levels and hence have more noticeable effects at small $q$ (Sec.~\ref{sec: energy levels}).
Meanwhile, FS effects are expected to be strong for $q<2q_F$~\cite{Marino2023,fetter2003quantum}.
Also, an analysis of the Fourier components of the densities (\cref{Table0,Table1}) highlights that higher-order harmonics are non-negligible, e.g. for $q = 2 q_{\rm int}$, which means the linear response theory does not solve the problem exactly.
The same tendency appears in \cref{chi01}, where $\chi_{01}$ is displayed, suggesting that the FS effects are controllable when predicting the perturbed isoscalar densities with isovector external fields.

In contrast to Figs.~\ref{chi00} and \ref{chi01}, the RPA and PBC results of $\chi_{10}$ and $\chi_{11}$ in Figs.~\ref{chi10} and \ref{chi11} only show a good agreement when $\beta = 0.94$ and $q_{\rm int}>5$.
For other isospin asymmetries, the FS effects are overall strong, although they slowly decrease as the transferred momentum increases.

Although calculated in the same system,
$\chi_{10}$ and $\chi_{11}$ show considerably larger deviations from the response functions in the TL 
than $\chi_{00}$ and $\chi_{01}$, 
indicating that the isovector density is more sensitive to the FS effects than the isoscalar density.

We speculate that the enhanced sensitivity of the isovector densities compared with the isoscalar densities arises from the specific behavior of the of the response functions (See also Fig.~\ref{beta0.50} and App.~\ref{chinpresponse} below).
In finite systems, the PBC introduces constraints analogous to those imposed by external fields, thereby contributing to the observed FS effects. 
Since the response functions $\chi_{nn}$ is overall negative and $\chi_{np}$ becomes positive around $q = 2\ \rm fm^{-1}$ , if, e.g., a positive external field is introduced in the neutron channel, it turns out that $\rho_n$ will decrease while $\rho_p$ will increase.
As a consequence, the change in the isovector densities, that is the neutron-proton density difference will be much larger than the sum, indicating a greater sensitivity of the isovector density.

\subsection{Response functions}
\label{RF}

\begin{figure*}[t!]
    \centering
    \includegraphics[width=0.49\linewidth]{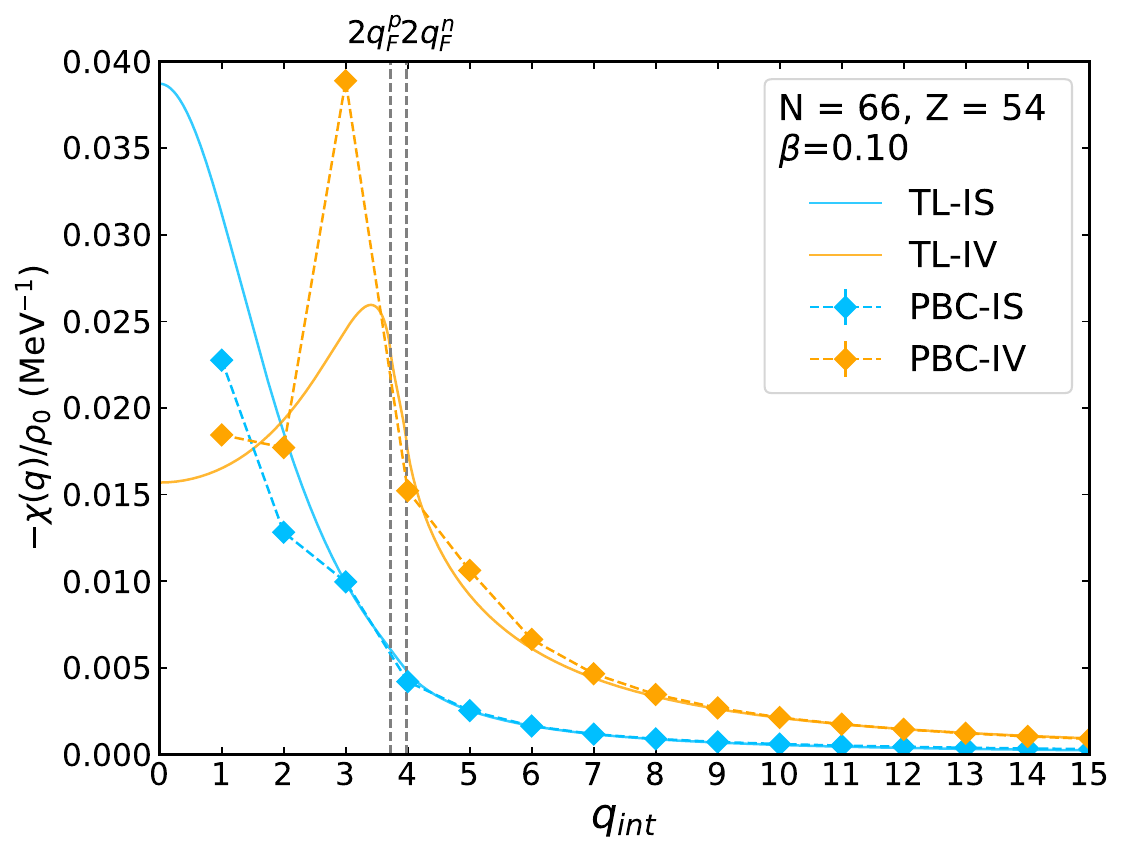}
    \includegraphics[width=0.49\linewidth]{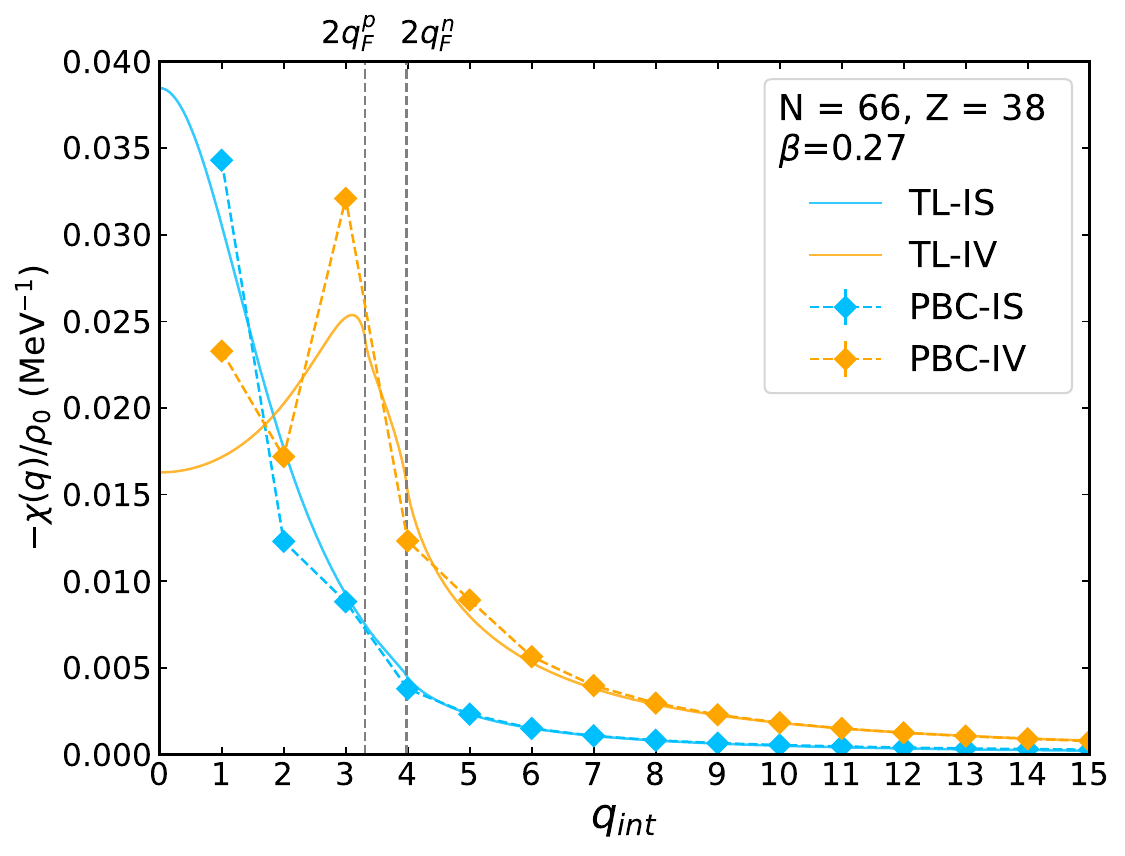}
    \includegraphics[width=0.49\linewidth]{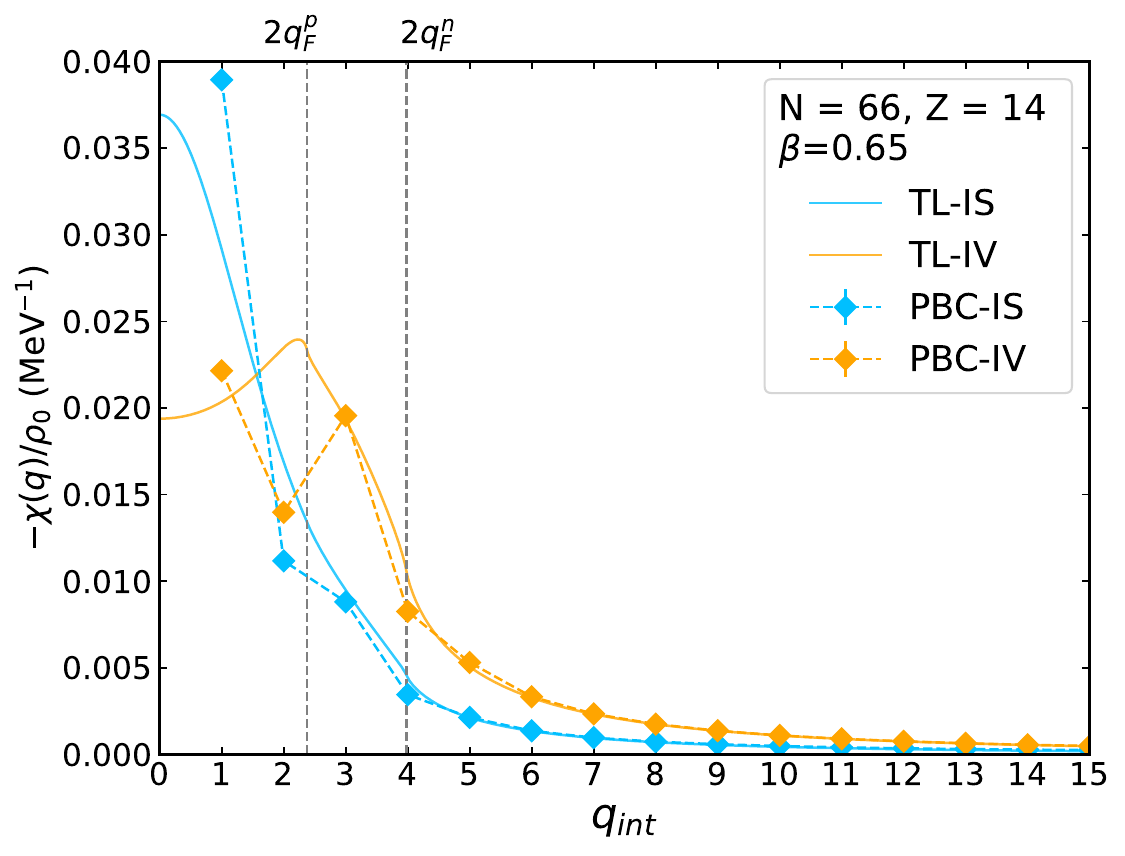}
    \includegraphics[width=0.49\linewidth]{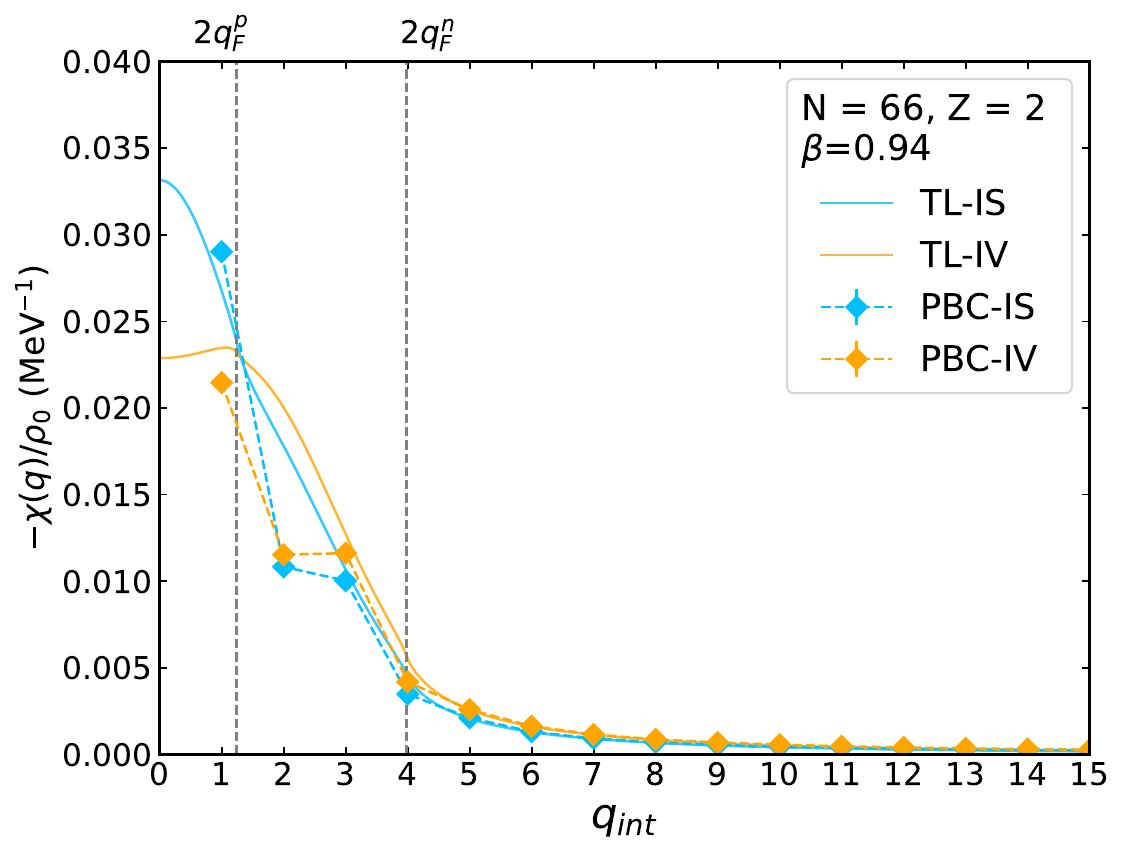}
    \caption{
    Response functions $-\chi (q) / \rho_0$ as a function of the transferred momentum $q = 2\pi/L q_{\rm int}$. Finite-$A$ calculations with 66 neutrons and different numbers of protons of the isoscalar (isovector) response are shown as dots (diamonds). For comparison, the corresponding RPA response functions in the TL are shown as solid (dash-dotted) lines. Vertical dashed lines denote $2q_F^N$ and $2q_F^P$, with $q_F^{N,P}$ being the neutron and proton Fermi momenta.}
    \label{EnRe}
\end{figure*}

The main results of our work consist of predictions of the static response functions in the finite-$A$ system.
While studying the ratio between densities and strengths is useful to understand the dependence on $v_q$, for an accurate determination of $\chi_{00}$ and $\chi_{11}$, we find it more convenient to extract the response functions by fitting perturbed energies using \cref{EnResponse}.
By using a least squares method, the response functions, as well as the fitting errors, are obtained.
As above, we fix $N=66$ and $\rho_0 = 0.16\ \rm fm^{-3}$ and vary $q$ and $v_q$ for different $\beta$'s.
In the following plots, we show $-\chi (q) / \rho_0$, which is a positive quantity.
Note that, while in principle we could mix up isoscalar and isovector fields in \cref{EnResponse} to probe the non-diagonal responses, in practice it is easier from a numerical point of view to consider either a purely isoscalar or isovector perturbation, thus focusing on the diagonal $\chi_{00}$ and $\chi_{11}$.

In \cref{EnRe}, the response functions obtained in the finite-$A$ systems are shown as blue and orange dots for the isoscalar and isovector channels, respectively.
For comparison, the corresponding RPA results in TL are shown as solid and dashed-dotted lines, respectively.
The fits to the energies are accurate, and fit uncertainties on $\chi(q)$ are typically negligible.
The calculated response function in the isoscalar channel agrees well with the results in TL when $q>2q_F$.
Numerical results with $q<2q_F$ show larger oscillations, which make them deviate more from the smooth TL functions, as already pointed out above.
In the isovector channel, a peak appears both in the response functions calculated numerically in the finite-$A$ system and in TL.
With increasing $\beta$, the peak moves to higher $q$ and its magnitude increases.

\begin{figure}[t]
\centering
\includegraphics[width = 8cm]{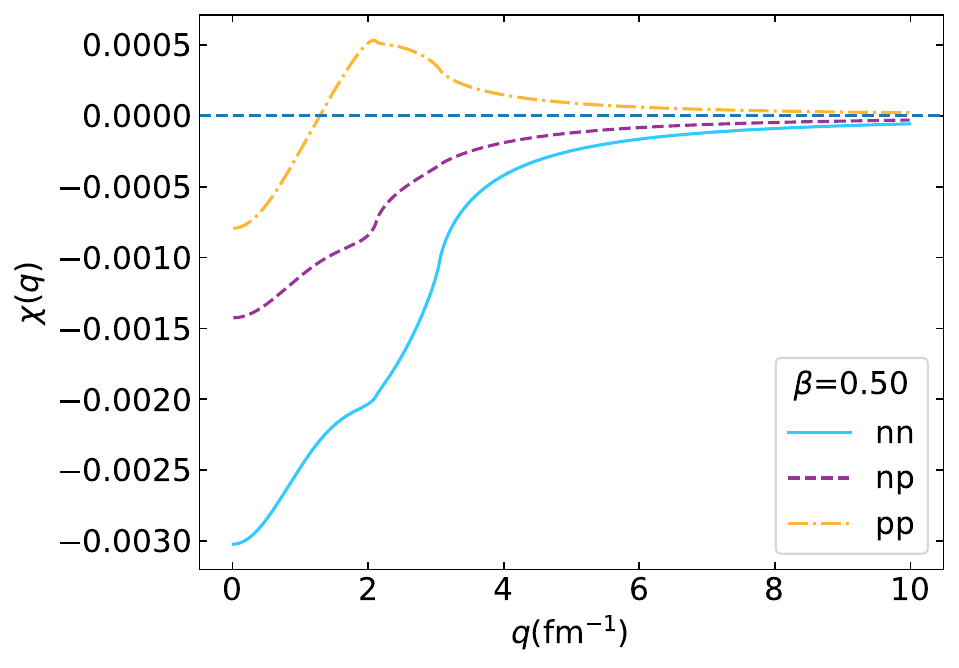}
\caption{Response functions calculated with RPA in the TL.
The response functions $\chi_{nn}$ and $\chi_{pp}$ are shown as blue and orange lines, respectively.
The response functions $\chi_{np}$ and $\chi_{np}$ are identical and shown as purple lines.
}
\label{beta0.50}.
\end{figure}

This peak is a somewhat peculiar feature of the isovector response, and does not appear either in the free gas response or in the isoscalar response of the interacting system.
To qualitatively understand its origin, we focus on the TL response in the neutron-neutron, proton-proton, and neutron-proton channels.
The response functions $\chi_{00}$ and $\chi_{11}$ are given by
\begin{align}
\chi_{00}(q) = \chi_{nn}(q) + \chi_{pp}(q) + 2 \chi_{np}(q),\label{ReIS}
\end{align}
\begin{align}
\chi_{11}(q) = \chi_{nn}(q) + \chi_{pp}(q) - 2 \chi_{np}(q),\label{ReIV}
\end{align}
where $\chi_{\tau\tau'}$ describes the response of the particles of type $\tau$ in the presence of an external fields acting on particles of type  $\tau'$, and we have used $\chi_{np}=\chi_{pn}$.
The response functions $\chi_{nn}(q)$, $\chi_{pp}(q)$, and $\chi_{np}(q)$ are shown in \cref{beta0.50} in the TL for $\beta = 0.50$.
The response functions $\chi_{nn}$ and $\chi_{pp}$ are negative and monotonically increasing.
(They differ in ANM, due to $N$ and $Z$ being different.)
This is the same qualitative behaviors as the free-gas Lindhard function, and reflects the behavior observed in both SNM and PNM, and in the ANM isoscalar response.
In contrast, the response function $\chi_{np}=\chi_{pn}$ is initially negative for low $q$, but changes sign at a relatively small momentum.
For $q \to \infty$, it converges to $0^+$,
(see App.~\ref{chinpresponse} for the discussion of the different behavior of $\chi_{nn}$ and $\chi_{np}$).
In short, the surface terms, which are associated with $q^2$, govern the relations between $\chi_{nn}$ and $\chi_{np}$ at $q\rightarrow \infty$.
Notably, the quantities $C^\Delta_0-C^\Delta_1$ and $C^\tau_0-C^\tau_1$ imply that $\chi_{nn}$ and $\chi_{np}$ exhibit different signs.

$\chi_{nn}$ and $\chi_{pp}$ have positive slope, while the derivative of $\chi_{np}$ changes its sign. 
As a result, the derivatives of $\chi_{00}(q)$ and $\chi_{11}(q)$ might also change sign. 
Namely, the response functions can show a peak, i.e., a local minimum or maximum.
Our numerical calculations suggest that this shows up as a minimum of $\chi_{11}(q)$ [maximum of $- \chi_{11}(q)/\rho_{0}$ in Fig.~\ref{EnRe}], while no peak  appears in $\chi_{00}(q)$.
Indeed, looking at Eq.~\eqref{ReIS} and Fig.~\ref{beta0.50}, the derivative $\chi_{00}^{\prime}(q)$ always remains positive, as the slope of $\chi_{nn}+\chi_{pp}$ is larger in magnitude than that of $2 \chi_{np}$.
In contrast, the minus sign in Eq.~\eqref{ReIV} leads to a competition between $\chi^{\prime}_{nn}+\chi^{\prime}_{pp} >0 $ and $-2 \chi_{np}^{\prime}<0$, the latter being large and possibly comparable in magnitude to the former for relatively small momenta.

The different behavior of $\chi_{nn}$, $\chi_{pp}$, and $\chi_{np}$ is still to be investigated.
At the HF level, the unperturbed response functions are diagonal in the isospin projections.
Therefore, $\chi_{nn}$, $\chi_{pp}$ are finite (and given by Lindhard functions). Thus, they are negative and scale as $\sim q^{-2}$ at large momenta.
In contrast, $\chi_{np}$ vanishes at the HF level, and becomes finite only when a coupling between $p$ and $n$ is induced, as e.g. by the residual $ph$ interaction in RPA.

Our argument helps interpreting the qualitative trend of the isovector response. However, we warn that position and magnitude of the peak are quite sensitive to the asymmetry $\beta$, as well as to the parameters of the EDF, in particular the density-gradient coefficients $C_{0}^\Delta$ and $C_{1}^\Delta$. 
More calculations are needed to study the sensitivities of the peak, which are important to the future studies on the isovector static response problems.

\section{Conclusions and perspectives}\label{Conclusion}
In this work, we have extended the finite-$A$ DFT method proposed in Ref.~\cite{Marino2023} to the case of isospin-asymmetric nuclear matter.
We have investigated the effect of external static perturbations in the different isospin channels on the properties of the system, namely, energies per particles, densities, and single-particle energy levels.
We have found that using $N = 186$ and $Z = 114$ is effective in minimizing the FS effects on the kinetic energy.

The isoscalar and isovector static response functions have been evaluated, and finite-size effects have been discussed in relation to RPA predictions in the thermodynamic limit.
FS errors tend to decrease as the momentum of the perturbation increases.
The response functions $\chi_{00}$ and $\chi_{01}$ in the finite-$A$ system typically match the RPA results well for $q/q_{min}>4$.
In contrast, FS effects $\chi_{10}$ and $\chi_{11}$, which are related to the impact of the external potentials on the isovector density of the system, remain large in most cases.
Investigating $\chi_{11}$, we have noticed that it features a peculiar non-monotonic behavior (in contrast to both the other DFT response functions and the free response), showing a peak at small $q/q_{min}$.

This study represents a step forward in the program started in Ref.~\cite{Marino2021} for constructing \textit{ab initio}-based nuclear EDFs.
Further constraints on the EDF surface terms will be determined by matching future \textit{ab initio} calculations of asymmetric nuclear matter to the DFT predictions obtained with our finite-$A$ approach.
As a next step, we plan to extend the formalism to address spin-polarized matter.

\section*{Acknowledgments}
F.M. acknowledges the CINECA awards AbINEF (HP10B3BG09) and RespGF (HP10BQMECT) under the ISCRA initiative, for the availability of high-performance computing resources and support.
This work is supported by the Deutsche Forschungsgemeinschaft (DFG) through Project-ID 279384907 - SFB 1245 and through the Cluster of Excellence “Precision Physics, Fundamental Interactions, and Structure of Matter” (PRISMA+ EXC 2118/1) funded by the DFG within the German Excellence Strategy (Project ID 39083149).

\appendix

\section{Kohn-Sham potentials}\label{Details}
In this work, we focus on the spin-saturated systems and the spin-orbit density is very small.
Therefore, the spin-density term is dropped in \cref{PotDensity} and we have
\begin{widetext}
\begin{align}
\begin{aligned}
\label{eq: energy density ez}
\mathcal{E}(z) = &\mathcal{E}_{\rm kin}(z) + \sum_{\gamma}(c_{\gamma,0}+c_{\gamma,1}\beta^{2})\rho^{\gamma+1} + \sum_{t} \Big[ C^\tau_t \rho_t(z)\tau_t(z) + C^{\Delta\rho}_t \rho_t(z)\Delta\rho_t(z) - C_t^{\nabla J} \rho_t \nabla \cdot \mathbf{J}_t (z)\Big) \Big].
\end{aligned}
\end{align}
Then, the potential $U_q(z)$ is calculated with the definition \cref{Def},
\begin{align}
\left\{
\begin{aligned}
U_q (z) = \frac{\delta E}{\delta \rho_q} = &\sum_{\gamma} \frac{(c_{\gamma,0}+c_{\gamma,1}\beta^{2})\rho_0^{\gamma+1}}{\delta \rho_0(z)} + \sum_{t} \left[ \frac{\delta (C^{\tau}_t \rho_t \tau_t)}{\delta \rho_q} + \frac{\delta (C^{\Delta\rho}_t \rho_t \Delta \rho_t)}{\delta \rho_q} + \frac{\delta (C_t^{\nabla J}\rho_t \nabla \cdot J_t)}{\delta \rho_q} \right]\\
=& \sum_\gamma[(\gamma+1)c_{\gamma,0}+2\beta(\tau_z-\beta)c_{\gamma,1}+(\gamma+1)c_{\gamma,1}\beta^2]\rho_0^\gamma\\
&+ C^{\tau}_0 \tau_0 +  \tau_z C^{\tau}_1 \tau_1 + 2C^{\Delta \rho}_0 \Delta \rho_0 + 2\tau_z C^{\Delta}_0 \Delta \rho_1 + C^{\nabla J}_0 \nabla \cdot \mathbf{J}_0 + \tau_z C^{\nabla J}_1 \nabla \cdot \mathbf{J}_1,
\end{aligned}
\right.
\label{Uq}
\end{align}
\end{widetext}
where $\tau_z = \pm 1$ for neutrons and protons, respectively. 

Moreover, the effective mass and spin-orbit potential are calculated as
\begin{align}
\frac{\hbar^2}{2m^*_q(z)} = \frac{\delta E}{\delta \tau_q} =  \frac{\hbar^2}{2m} + C_0^\tau \rho_0(z) + \tau_z C_1^\tau \rho_1(z),\label{mq}
\end{align}
\begin{align}
W_q(z) = \frac{\delta E}{\delta \mathbf{J}_n} = -(C^{\nabla J}_0 \nabla \rho_0(z) + \tau_z C^{\nabla J}_1 \nabla \rho_1(z)),\label{Wq}
\end{align}
where we used the relation
\begin{align}
\rho ( \nabla \cdot \mathbf{J}) = \nabla \cdot (\rho \mathbf{J}) -  (\nabla \rho) \cdot \mathbf{J}.\label{divergence}
\end{align}
The first term at r.h.s turns out to be zero considering the boundary condition.

\section{Response function calculated with RPA}\label{RPAresponse}
In RPA, the calculations of the response functions are based on the Bethe-Salpeter equation for the correlated propagator 
\cite{PASTORE20151}
\begin{widetext}
\begin{align}
\begin{aligned}G_{\rm RPA}^{(\alpha)}(\mathbf{k}_1,\mathbf{q},\omega)=&G_{\rm HF}(\mathbf{k}_1,\mathbf{q},\omega)\\&+G_{\rm HF}(\mathbf{k}_1,\mathbf{q},\omega)\sum_{(\alpha^{\prime})}\int\frac{d^3\mathbf{k}_2}{(2\pi)^3}V_{ph}^{(\alpha,\alpha^{\prime})}(\mathbf{k}_1,\mathbf{k}_2)G_{\rm RPA}^{(\alpha^{\prime})}(\mathbf{k}_2,\mathbf{q},\omega),\end{aligned}\label{BSequation}
\end{align}
\end{widetext}
where $G_{\rm RPA}^{(\alpha)}$ is the RPA Green's function and $G_{\rm HF}$ is the Green's function in the Hartree-Fock approximation.
$(\alpha) = (\tau,\tau', S, M)$ are the quantum numbers associated with the interaction channel.
In this work, the quantum numbers $S$ and $M$ are dropped since we focus on spin-saturated systems and $S=M=0$.
$\mathbf{k}_1$ is the hole momentum of the particle-hole pairs with the quantum number $(\alpha)$ and 
\begin{align}
\label{eq: chi and Grpa}
\chi_{\rm RPA}^{(\alpha)}(\mathbf{q},\omega) = n_d \int\frac{d^{3}\mathbf{k_1}}{(2\pi)^{3}}G_{\rm RPA}^{(\alpha)}(\mathbf{k_1},\mathbf{q},\omega),
\end{align}
where $n_d = 2$ is the spin-degeneracy and $\chi_{\rm RPA}^{(\alpha)}(\mathbf{q},\omega)$ is the RPA response function in the TL.
Moreover, the interaction $V_{ph}^{(\alpha,\alpha^{\prime})}$ is called the particle-hole residual interaction, and is obtained from the second functional derivative with respect to densities taken at the level of the Hartree-Fock solution \cite{10.1119/1.1975177,RevModPhys.43.1}.

The Bethe-Salpeter equation is solved in each channel $(\alpha)$.
The method of solving the Bethe-Salpeter equation is given in \cite{GARCIARECIO1992293,PhysRevC.80.024314,PhysRevC.100.064301}.
In short, a set of algebraic equations are obtained by multiplying the Bethe-Salpeter equation with the factors $1, k^2, kY_{1,0}, k^2|Y_{1,\pm 1}|$, and $k^2|Y_{1,0}|^2$ and integrating over the momentum $\mathbf{k}_1$. 
Because of the properties of the \textit{ph} interaction, the propagator $G^{(\alpha)}_{\rm RPA}$ only enters the equations as $\langle G^{(\alpha)}_{\rm RPA} \rangle, \langle k^2 G^{(\alpha)}_{\rm RPA} \rangle, \langle kY_{10}G^{(\alpha)}_{\rm RPA} \rangle, \langle k^2|Y_{1, \pm 1}|^2 G^{(\alpha)}_{\rm RPA} \rangle$, and $\langle k^2|Y_{1, 0}|^2 G^{(\alpha)}_{\rm RPA} \rangle$. 
Then, a complete set of algebraic equations in such variables are obtained and the RPA propagator $\langle G_{\rm RPA}^{\alpha}\rangle$ can be found.
The response function for the ANM with $S=M=0$ has been obtained in Ref.~\cite{PhysRevC.100.064301}, with the spin-orbit terms included.

\section{Change of energy levels}\label{Energychange}
In this section, we will show that the energy levels in perturbed systems remain approximately the same as in  the unperturbed systems when the perturbation has large transferred momentum $q$.

Let us consider the expectation value of the Hamiltonian in the KS-DFT equation:
\begin{align}
    \begin{aligned}
        E = &\langle \hat H \rangle \\
        = &\langle \psi(\mathbf x)| -\nabla\cdot\frac{\hbar^2}{2m_q^*(\mathbf{x})}\nabla+U_q(\mathbf{x})\\
        &+v_q(\mathbf{x})+\mathbf{W}_q(\mathbf{x})\cdot(-i)\left(\nabla\times\boldsymbol{\sigma}\right)|\psi(\mathbf x)\rangle.
    \end{aligned}
    \label{energyvalue}
\end{align}
In the unperturbed state, translational invariance holds and the field terms $\frac{\hbar^2}{2m_q^*(\mathbf{x})}$, $U_q(\mathbf{x})$, and $\mathbf{W}_q(\mathbf{x})$ are constants, denoted as $\frac{\hbar^2}{2m_{q,0}^*}$, $U_{q,0}$, and $\mathbf{W}_{q,0}$
Therefore, the eigenvalue reads
\begin{align}
    E = \int-\frac{\hbar^2}{2m_{q,0}^*}\psi^*(z)\psi'' + (U_{q,0} + W_{q,0} K_{n_xn_y}) \psi^*(z)\psi(z) dz,
\end{align}
where $q = n,p$ for neutrons and protons, respectively.

In the presence of an external field with the form 
\begin{align}
\left\{ 
\begin{aligned}
&v_0(z)=2v_q\cos{(q/q_{\rm min} \frac{2\pi}{L}z)},\\
&v_1(z)=2v_q\cos{(q/q_{\rm min} \frac{2\pi}{L}z)}\hat{\tau_z},
\end{aligned}
\right.
\end{align}
all the densities terms turn to cosine functions oscillating around the original value (at linear order):
\begin{align}
&\rho_t(z) \approx \rho_{t,0} + A_{\rho_t} \cos(q/q_{\rm min} \frac{2\pi}{L}z),\\
&\tau_t(z) \approx \tau_{t,0} + A_{\tau_t} \cos(q/q_{\rm min} \frac{2\pi}{L}z),\\
&\mathbf{J}_{t}(z) \approx \mathbf{J}_{t,0} + A_{\mathbf{J}_t} \cos(q/q_{\rm min} \frac{2\pi}{L}z),
\end{align}
where $A_{\rho_t}$, $A_{\tau_t}$, and $A_{\mathbf{J}_t}$ are constants determined by the static response theory.
Therefore, the field terms are calculated with \cref{Uq,mq,Wq} and then have the form:
\begin{align}
&\frac{\hbar^2}{2m_q^*(\mathbf{x})} \approx \frac{\hbar^2}{2m_{q,0}^*}+ A_{q,1}\cos(q/q_{\rm min} \frac{2\pi}{L}z)\\
&\begin{aligned}
U_q(z) \approx & U_{q,0} + \sum_{\gamma} A_{\gamma} \cos^{\gamma}(q/q_{\rm min} \frac{2\pi}{L}z) \\
&+ A_{q,2} \cos(q/q_{\rm min} \frac{2\pi}{L}z) + A_{q,3} \sin(q/q_{\rm min} \frac{2\pi}{L}z)
\end{aligned}\\
&W_q(z) \approx W_{q,0} + A_{q,4}\sin(q/q_{\rm min} \frac{2\pi}{L}z),
\end{align}
where $A$ are constants.
Hence, the eigenvalue can be calculated as 
\begin{align}
\begin{aligned}
    E \approx &\int dz -\frac{\hbar^2}{2m_{q,0}^*}\psi^*(z)\psi''(z)\\
    &+ (U_{q,0} + W_{q,0} K_{n_xn_y}) \psi^*(z)\psi(z),\\
    &+ q/q_{\rm min} \frac{2\pi}{L}A_{q,1}\psi^*(z)\sin(q/q_{\rm min} \frac{2\pi}{L}z)\psi'(z)\\
    &- A_{q,1}\psi^*(z)\cos(q/q_{\rm min} \frac{2\pi}{L}z)\psi''(z)\\
    &+\psi^*(z)\Big[\sum_{\gamma} A_{\gamma} \cos^\gamma(q/q_{\rm min} \frac{2\pi}{L}z)\\
    &+ A_{q,2} \cos(q/q_{\rm min} \frac{2\pi}{L}z) \\
    &+ (A_{q,3} + A_{q,4})\sin(q/q_{\rm min} \frac{2\pi}{L}z)\Big]\psi(z).
\end{aligned}
\label{eigenvalue}
\end{align}
When the the transferred momentum $q/q_{\rm min}$ is sufficiently large, i.e., the frequency of the $\cos(q/q_{\rm min} \frac{2\pi}{L}z)$ and $\sin(q/q_{\rm min} \frac{2\pi}{L}z)$ is much larger than the frequency of the wavefunctions $\psi(z)$, the contribution to the integration over $z$ from the second to the fifth line in \cref{eigenvalue} is negligible and the integrals \cref{eigenvalue} and \cref{energyvalue} become the same.
Therefore, the energy levels of the systems remains approximately unchanged with large transferred momentum $q/q_{\rm min}$.

\section{The $\chi_{np}$ response}\label{chinpresponse}
By multiplying the BS equation \cref{BSequation} with the factors $1, k^2, kY_{1,0}, k^2|Y_{1,\pm 1}|$, and $k^2|Y_{1,0}|^2$, and integrating over the momentum $\mathbf{k}_1$, the equation is recast into a matrix equation \cite{Davesne2014}:

\begin{align}
\begin{pmatrix}A_{nn}&A_{np}\\A_{pn}&A_{pp}\end{pmatrix}\begin{pmatrix}X_{nn}\\X_{pn}\end{pmatrix}=\begin{pmatrix}B_{n}\\0\end{pmatrix},\label{AXB}
\end{align}
where
\begin{widetext}
\begin{align}
\begin{aligned}
&X_{nn}=\begin{pmatrix}\langle G_{\rm RPA}^{(nn)}\rangle\\\langle k^2G_{\rm RPA}^{(nn)}\rangle\\\sqrt{\frac{4\pi}3}\langle kY_{10}G_{\rm RPA}^{(nn)}\rangle\end{pmatrix},\quad X_{pn}=\begin{pmatrix}\langle G_{\rm RPA}^{(pn)}\rangle\\\langle k^2G_{\rm RPA}^{(pn)}\rangle\\\sqrt{\frac{4\pi}3}\langle kY_{10}G_{\rm RPA}^{(pn)}\rangle\}\end{pmatrix},\quad B_n=\begin{pmatrix}\beta_0^{(n)}\\q^2\beta_2^{(n)}\\q\beta_1^{(n)}\end{pmatrix},\\
&A_{nn}=\begin{pmatrix}1-\beta_0^{(n)}\widetilde{W}_1^{(nn,0)}-q^2\beta_2^{(n)}W_2^{(nn,0)}&-\beta_0^{(n)}W_2^{(nn,0)}&2q\beta_1^{(n)}W_2^{(nn,0)}\\-q^2\beta_2^{(n)}\widetilde{W}_1^{(nn,0)}-q^4\beta_5^{(n)}W_2^{(nn,0)}&1-q^2\beta_2^{(n)}W_2^{(nn,0)}&2q^3\beta_4^{(n)}W_2^{(nn,0)}\\-q\beta_1^{(n)}\widetilde{W}_1^{(nn,0)}-q^3\beta_4^{(n)}W_2^{(nn,0)}&-q\beta_1^{(n)}W_2^{(nn,0)}&1+2q^2\beta_3^{(n)}W_2^{(nn,0)}\end{pmatrix},\\
&A_{np}=\begin{pmatrix}-\beta_0^{(n)}\widetilde{W}_1^{(np,0)}-q^2\beta_2^{(n)}W_2^{(np,0)}&-\beta_0^{(n)}W_2^{(np,0)}&2q\beta_1^{(n)}W_2^{(np,0)}\\-q^2\beta_2^{(n)}\widetilde{W}_1^{(np,0)}-q^4\beta_5^{(n)}W_2^{(np,0)}&-q^2\beta_2^{(n)}W_2^{(np,0)}&2q^3\beta_4^{(n)}W_2^{(np,0)}\\-q\beta_1^{(n)}\widetilde{W}_1^{(np,0)}-q^3\beta_4^{(n)}W_2^{(np,0)}&-q\beta_1^{(n)}W_2^{(np,0)}&2q^2\beta_3^{(n)}W_2^{(np,0)}\end{pmatrix}
\end{aligned}\label{BSmatrix}
\end{align}
\end{widetext}
(see Ref.~\cite{Davesne2014} for the details of the functions and coefficients).
The matrices $A_{pp}$ and $A_{pn}$ are calculated by a simple swap of the labels $n\rightarrow p$.

\begin{figure}[h!]
\centering
\includegraphics[width = 8cm]{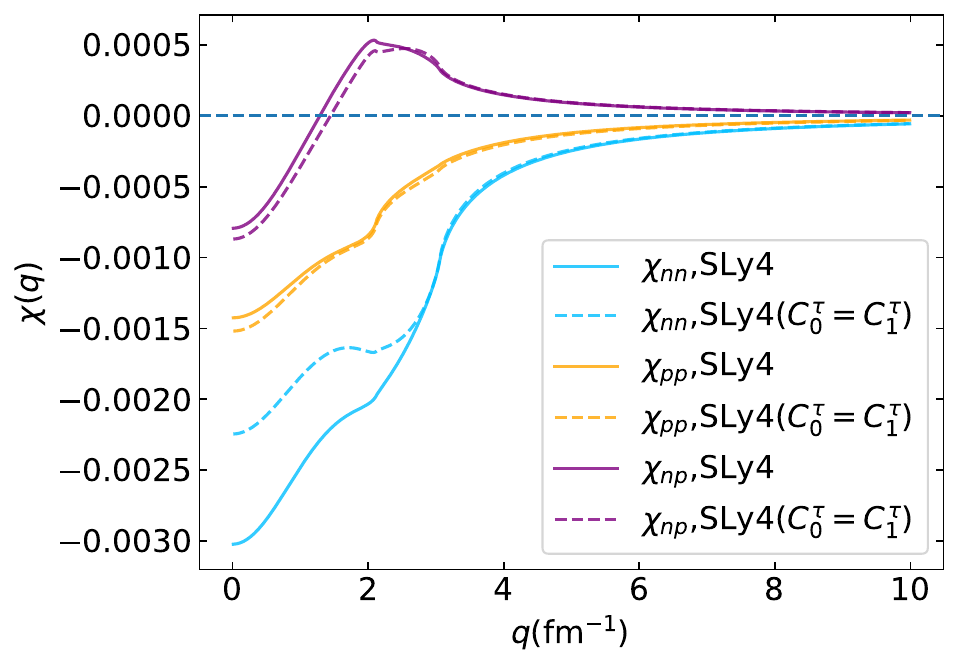}
\caption{Response functions at $\beta = 0.50$ calculated with RPA in the TL.
The response functions $\chi_{np}$, $\chi_{pp}$, and  $\chi_{np}$ calculated with the SLy4 EDF are shown as blue, orange, and purple solid lines, respectively.
The corresponding response functions calculated with SLy4 EDF but $C_0^\tau = C_1^\tau$ are shown as dashed lines.
}
\label{Change}.
\end{figure}

We now simplify the discussion of the general properties of the response by taking $C_0^\tau = C_1^\tau$, but leaving the remaining terms unchanged.
This assumption has no qualitative impact (see \cref{Change}).

Then, the matrix elements $A_{np}$ in \cref{BSmatrix} turn out to reduce to
\begin{align}
A_{pn}=\begin{pmatrix}-\beta_0^{(p)}\widetilde{W}_1^{(pn,0)}&0&0\\-q^2\beta_2^{(p)}\widetilde{W}_1^{(pn,0)}&0&0\\-q\beta_1^{(p)}\widetilde{W}_1^{(pn,0)}&0&0\end{pmatrix}.
\end{align}
The first line of $A_{pn}X_{nn} + A_{pp}X_{pn}$ gives
\begin{widetext}
\begin{align}
\begin{aligned}
\beta_0^{(p)}\widetilde{W}_1^{(pn,0)} \langle G_{\rm RPA}^{nn} \rangle = &(1-\beta_0^{(p)}\widetilde{W}_1^{(pp,0)}-q^2\beta_2^{(p)}W_2^{(pp,0)}) \langle G_{\rm RPA}^{pn} \rangle  \\
&- \beta_0^{(p)}W_2^{(pp,0)}\langle k^2G_{\rm RPA}^{pn} \rangle + 2q\beta_1^{(p)}W_2^{(pp,0)} \langle kY_{10}G_{\rm RPA}^{(pn)}\rangle .
\end{aligned}
\end{align}
\end{widetext}

We notice that the product $-\beta_0^{(p)}\widetilde{W}_1^{(pp,0)}\langle G_{\rm RPA}^{pn} \rangle$ dominates the r.h.s. at $q \rightarrow \infty$, because it scales as $q^2$.
(The functions $\beta_0^{(p)}$, $q\beta_1^{(p)}$, and $q^2\beta_2^{(p)}$ have the same orders of $q^0$ while $W_2^{(pp)}$ is a constant. See Ref.~\cite{Davesne2019} for the details of $\beta_i^\tau$).

Therefore, in the asymptotic limit
\begin{align}
   \widetilde{W}_1^{(pn,0)} \langle G_{\rm RPA}^{nn} \rangle \approx
    \widetilde{W}_1^{(pp,0)}\langle G_{\rm RPA}^{pn} \rangle,
\end{align}
and, reminding Eq.~\eqref{eq: chi and Grpa}, the relation between the response functions $\chi_{nn}$ and $\chi_{pp}$ is determined by the ratio $\tilde W_1^{(pn,0)}/\tilde W_1^{(pp,0)}$.
Also, note that $\tilde W_1^{(\tau\tau^\prime,0)}$ contains a spin-orbit correction to the original coefficients  $W_1^{(\tau\tau^\prime,0)}$, which however is also negligible at $q \rightarrow \infty$.
In fact, the relations between response functions $\chi_{nn}$ and $\chi_{np}$ are insensitive to $C_t^{\nabla J}$.
The coefficients $W_1^{(\tau\tau^\prime,0)}$ are calculated as
\begin{align}
&W_1^{(pp,0)} = W_1^{(0,0)} + W_1^{(0,1)} - 8C_1^{\rho,\gamma}\rho_1\gamma\rho^{\gamma-1},\\
&W_1^{(p,n,0)} = W_1^{(0,0)} - W_1^{(0,1)},
\end{align}
\begin{align}
\begin{aligned}
\frac{1}{4}W_{1}^{(0,0)} =&2C_0^{\rho0}+(2+\gamma)(1+\gamma)C_0^{\rho,\gamma}\rho_0^\gamma\\
&+\gamma(\gamma-1)  \times C_{1}^{\rho,\gamma}\rho_{0}^{\gamma-2}\rho_{1}^{2}\\
&-\left[2C_{0}^{\Delta}+\frac{1}{2}C_{0}^{\tau}\right]q^{2}, \label{W100} 
\end{aligned}
\end{align}
\begin{align}
&\frac14W_1^{(0,1)} =2C_1^{\rho0}+2C_1^{\rho,\gamma}\rho_0^{\gamma}-\left[2C_1^{\Delta}+\frac12C_1^{\tau}\right]q^2. \label{W101}
\end{align}
At $q \to \infty$, the bulk terms are negligible and the sign of $\frac{\tilde W_1^{(pn,0)}}{\tilde W_1^{(pp,0)}}$ depends on the quantity 
\begin{align}
   \frac{8(C_0^\Delta - C_1^\Delta) + 2(C_0^\tau - C_1^\tau)}{8(C_0^\Delta + C_1^\Delta) + 2(C_0^\tau + C_1^\tau)},
\end{align}
which is found to be positive in several widely-used EDFs, e.g., the SLy4, SkM, and SkI3 EDFs (see~\cite{Marino2023}).
Hence, we conclude that the response functions $\chi_{nn}$ and $\chi_{np}$ have opposite signs at $q\rightarrow \infty$.


\end{document}